\documentclass{CSML}

\def\dOi{12(3:5)2016}
\lmcsheading%
{\dOi}
{1--28}
{}
{}
{Jul.~31, 2015}
{Sep.~\phantom05, 2016}
{}

\ACMCCS{[{\bf Theory of computation}]: Logic---Automated reasoning\,/\,Constraint and logic programming}

\usepackage{hyperref}
\usepackage{url}
\usepackage{amsfonts}
\usepackage{amsmath}
\usepackage{amssymb}
\usepackage{amsthm}
\usepackage{multirow}
\usepackage{graphicx}
\usepackage[labelfont=bf]{caption}
\usepackage{float}
\usepackage[boxed]{algorithm}
\usepackage[noend]{algorithmic}

\newcommand{\X}{\mathbf{X}}
\newcommand{\D}{\mathbf{D}}
\newcommand{\C}{\mathbf{C}}
\newcommand{\alldiff}{\texttt{alldifferent} }

\renewcommand{\paragraph}[1]{\noindent{}\emph{#1}}

\begin{document}

\title[Solving finite-domain linear constraints in presence of the \texttt{alldifferent}]
      {Solving finite-domain linear constraints in presence of the \texttt{alldifferent}}
\author[M.~Bankovi\' c]{Milan Bankovi\' c}
\address{Faculty of Mathematics, University of Belgrade\\ Studentski Trg 16, 11000 Belgrade, Serbia}
\email{milan@matf.bg.ac.rs}


\begin{abstract}
\noindent In this paper, we investigate the possibility of improvement of the widely-used filtering algorithm for the linear constraints in constraint satisfaction problems in the presence of the alldifferent constraints. 
In many cases, the fact that the variables in a linear constraint are also constrained by some alldifferent constraints
may help us to calculate stronger bounds of the variables, leading to a stronger constraint propagation. We propose an improved filtering algorithm that targets such cases. We provide a detailed description
of the proposed algorithm and prove its correctness. We evaluate the approach on five different problems that involve combinations of the linear 
and the alldifferent constraints. We also compare our algorithm to other relevant approaches. The experimental results show a great potential of the proposed improvement.
\end{abstract}

\maketitle

\section{Introduction}
\label{sec:intro}

A constraint satisfaction problem (CSP) over a finite set of variables is the problem of finding values for the 
variables from their finite domains such that all the imposed constraints are satisfied. There are many practical
problems that can be expressed as CSPs, varying from puzzle solving, scheduling, combinatorial design problems,
and so on. Because of the applicability of CSPs, many different solving techniques have been considered in the 
past decades. More details about solving CSPs can be found in \cite{csp_handbook}.

A special attention in CSP solving is payed to so-called \emph{global constraints} which usually have their own specific
semantics and are best handled by specialized \emph{filtering} algorithms that remove the values inconsistent with the constraints and trigger the constraint propagation. These filtering algorithms usually consider
each global constraint separately, i.e.~a filtering algorithm is typically executed on a constraint without any awareness of the existence of other global constraints. In some cases, however, it would be beneficial to take into account the presence of other global constraints in a particular CSP, since this could lead to a stronger propagation. For example, many common problems can be modeled by CSPs that include some combination of the \alldiff constraints (that constrain their
variables to take pairwise distinct values) and \emph{linear} constraints that are relations of the form
$a_1 \cdot x_1 + \ldots{} + a_n \cdot x_n \le c$ ($\le$ can be replaced by some other relation symbol, such as $\ge$, $=$, etc.). A commonly used filtering algorithm for linear constraints (\cite{stuckey_bounds}) enforces \emph{bound consistency} on a constraint ---
for each variable the maximum and/or the minimum are calculated, and the values outside of this interval are \emph{pruned} (removed) from the domain. The maximum/minimum for a variable is calculated based on the current maximums/minimums of other 
variables in the constraint. If we knew that some of the variables in the linear constraint were also constrained
by some \alldiff constraint to be pairwise distinct, we would potentially be able to calculate stronger bounds, leading
to more prunings. Let as consider the following example.

\begin{exa}
\label{ex:kakuro}
Consider an instance of the well-known \emph{Kakuro puzzle} (Figure \ref{fig:kakuro_ex}). Each empty cell should be
filled with a number from $1$ to $9$ such that all numbers in each line (a vertical or horizontal sequence of 
adjacent empty cells) are pairwise distinct and their sum is equal to the adjacent number given in the grid. The problem
is modeled by the CSP where a variable with the domain $\{1,\ldots{},9\}$ is assigned to each empty cell, variables
in each line are constrained with an \alldiff constraint and with one linear constraint that constrains the variables
to sum up to the given number. 

\begin{figure}[!h]
\begin{center}
  \includegraphics[width=150pt]{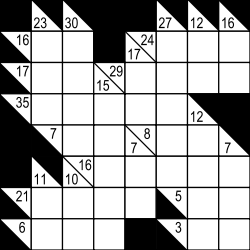}
\end{center}
\caption{Kakuro puzzle}
\label{fig:kakuro_ex}
\end{figure}

Consider, for instance, the first horizontal line in the bottom row --- it consists of three
cells whose values should sum up to $6$. Assume that the variables assigned to these cells are $x$, $y$ and $z$, respectively. These variables are constrained by the constraint $\mathtt{alldifferent}(x,y,z)$ and by the linear
constraint $x + y + z = 6$.
If the filtering algorithm for the linear constraint is completely unaware of the existence of the \alldiff constraint, it will deduce that
the values for $x$, $y$ and $z$ must belong to the interval $[ 1, 4 ]$, since a value greater than $4$ for the variable $x$
together with the minimal possible values for $y$ and $z$ (and that is $1$) sum up to at least $7$ (the symmetric situation is with variables $y$ and $z$). But if the filtering algorithm was aware of the presence of the \alldiff 
constraint, it would deduce that the feasible interval is $[1, 3]$, since the value $4$ for $x$ imposes that the 
values for $y$ and $z$ must both be $1$, and this is not possible, since $y$ and $z$ must be distinct. 
\end{exa}

The above example demonstrates the case when a modification of a filtering algorithm that makes it aware of the presence of other
constraints imposed on its variables may lead to a stronger propagation. In this paper, we consider this specific
case of improving the filtering algorithm for linear constraints in the presence
of the \alldiff constraints. The goal of this study is to evaluate
the effect of such improvement on the overall solving process. We designed and implemented a simple algorithm that 
strengthens the calculated bounds based on the \alldiff constraints. The paper also contains the proof of 
the algorithm's correctness as well as an experimental evaluation that shows a very good behaviour of our improved algorithm on specific problems that include a lot of \texttt{alldifferent}-constrained linear sums. We must stress that, unlike some other approaches, our algorithm does not establish bound consistency on a conjunction of a linear and an \alldiff constraint, but despite the weaker consistency level, it performs very well in practice. On the other hand, the main advantage of our algorithm is its 
generality, since it permits arbitrary combinations of the linear and the \alldiff constraints (in particular, the algorithm can handle the combination of a linear constraint with multiple \alldiff constraints that partially overlap with the linear constraint's variables).  

\section{Background}
\label{sec:background}

A Constraint Satisfaction Problem (CSP) is represented by a triplet $(\X, \D, \C)$, where
$\X = (x_1, x_2, \ldots{}, x_n)$ is a finite set of variables, $\D = (D_{x_1}, D_{x_2}, \ldots{}, D_{x_n})$ is a set of finite domains, where $D_{x_i}$ is the domain of the variable 
 $x_i$, $\C = \{C_1, C_2, \ldots{}, C_m\}$ is a finite set of constraints. A constraint $C \in \C$ over variables $x_{i_1}, x_{i_2}, \ldots{}, x_{i_k}$ is some subset of $D_{x_{i_1}} \times D_{x_{i_2}} \times \ldots{} \times D_{x_{i_k}}$. The number $k$ is called {\em arity} of the constraint $C$. 
A \emph{solution} of CSP is any $n$-tuple $(d_1, d_2, \ldots{}, d_n)$ from $D_{x_1} \times D_{x_2} \times \ldots{} \times D_{x_n}$ such that for each constraint  $C \in \C$ over variables  $x_{i_1}, x_{i_2}, \ldots{}, x_{i_k}$ $k$-tuple  $(d_{i_1}, d_{i_2}, \ldots{}, d_{i_k})$ is in $C$. A CSP problem is \emph{consistent} if it has a solution, and \emph{inconsistent} otherwise. Two CSP problems $P_1$ and $P_2$ are \emph{equivalent} if each solution of $P_1$ is
also a solution of $P_2$ and vice-versa. 

A constraint $C$ over variables $x_{i_1}, x_{i_2}, \ldots{}, x_{i_k}$ is \emph{hyper-arc consistent} if for each value $d_{i_r} \in
D_{x_{i_r}}$ ($r \in \{1,\ldots{},k\}$) there are values $d_{i_s} \in D_{x_{i_s}}$ for each $s \in \{1,\ldots{},k\} \setminus \{r\}$,
such that $(d_{i_1},\ldots{},d_{i_k}) \in C$. Assuming that the domains are ordered, a constraint $C$ over variables $x_{i_1}, x_{i_2}, \ldots{}, x_{i_k}$ is \emph{bound 
consistent} if for each value $d_{i_r} \in
\{min(D_{x_{i_r}}), max(D_{x_{i_r}})\}$ ($r \in \{1,\ldots{},k\}$) there are values $d_{i_s} \in [min(D_{x_{i_s}}),max(D_{x_{i_s}})]$
for each $s \in \{1,\ldots{},k\} \setminus \{r\}$, such that 
$(d_{i_1},\ldots{},d_{i_k}) \in C$. A CSP problem is \emph{hyper-arc consistent} (\emph{bound consistent}) if all its constraints are.

Constraints whose arity is greater then two are often called \emph{global} constraints. There are two types of global
constraints that are specially interesting for us here. One is the \alldiff constraint defined as follows:
$$
\mathtt{alldifferent}(x_{i_1}, x_{i_2}, \ldots{}, x_{i_k}) = \{ (d_{i_1}, d_{i_2}, \ldots{}, d_{i_k})\  |\  d_{i_j} \in D_{x_{i_j}}, 
r \neq s \Rightarrow d_{i_r} \neq d_{i_s}  \}
$$
The consistency check for the \alldiff constraint is usually reduced to the maximal matching problem in \emph{bipartite 
graphs} (\cite{hoeve}). The hyper-arc consistency on an \alldiff constraint can be enforced by Regin's algorithm (\cite{regin}). 

Another type of constraint interesting for us is the \emph{linear constraint} of the form:

$$
a_1 \cdot x_1 + \ldots{} + a_k \cdot x_k \Join c
$$
where $\Join \in \{ =, \ne, \le, <, \ge, > \}$, $a_i$ and $c$ are integers and $x_i$ are finite 
domain integer variables. Notice that we can assume without lost of generality that 
the only relations that appear in the problem are $\le$ and $\ge$. Indeed, a strict inequality $e < c$ ($e > c$) may always be replaced by $e \le c - 1$ ($e \ge c + 1$), an equality $e = c$ may be replaced by the conjunction $e \le c \wedge e \ge c$ and a disequality $e \ne c$ may be replaced by the disjunction $e \le c - 1 \vee e \ge c + 1$. These replacements
can be done in the preprocessing stage. The bound consistency on a linear constraint can be enforced by the filtering algorithm given, for instance, in \cite{stuckey_bounds}, and discussed in more details later in the paper. 

The CSP solving usually combines \emph{search} with \emph{constraint propagation}. By \emph{search} we mean dividing
the problem $P$ into two or more subproblems $P_1,\ldots{},P_n$ such that the solution set of $P$ is the union of
the solution sets of the subproblems $P_1,\ldots{},P_n$. The subproblems are then recursively checked for consistency one by one -- if any of them is consistent, the problem $P$ is consistent, too. The usual way to split the problem $P$ is to
consider different values for some variable $x$. On the other hand, the \emph{constraint propagation} uses 
inference to transform the problem $P$ into a simpler but equivalent problem $P'$. This is usually done by \emph{pruning}, i.e. removing the values from the variable domains that are found to be inconsistent with some of
the constraints and \emph{propagating} these value removals to other interested constraints. More detailed information
on different search and propagation techniques and algorithms can be found in \cite{csp_handbook}. 

\section{Linear constraints and \texttt{alldifferent}}
\label{sec:main}

In this section we consider the improvement of the standard filtering algorithm\footnote{The \emph{standard filtering algorithm} may not be the term that is typically used for this method in the literature, but it \emph{is} the standardly used method for calculating bounds in the linear constraints. We use the term \emph{standard filtering algorithm} to distinguish it from 
our \emph{improved filtering algorithm} for the linear constraints throughout the paper.} (\cite{stuckey_bounds}) for 
the linear constraints which takes into account the presence of the \texttt{alldifferent} constraints. We first briefly explain the standard algorithm, and then we discuss the improved algorithm and prove its correctness.

\subsection{Standard filtering algorithm} 
\label{subsec:standard}

Assume that we have the linear constraint $e \le c$, where $e \equiv a_1 \cdot x_1 + \ldots{} + a_n \cdot x_n$. The procedure $\mathtt{calculateBoundsStandard}(e, c, bounds)$ (Algorithm \ref{algo:bounds_standard}) implements the standard filtering algorithm. It receives $e$ and $c$ as inputs and returns $true$ if the constraint $e \le c$ is consistent, and $false$ otherwise. It also has an output parameter $bounds$ to which it stores 
the calculated bounds (only if the constraint is consistent). The idea is to first calculate the minimum of the left-hand side expression $e$ (denoted by $min(e)$) in the following way:
$$
min(e) = min(a_1 \cdot x_1) + \ldots{} + min(a_n \cdot x_n)
$$
where 
$$
\begin{array}{ccc}
min(a_i \cdot x_i) & = & \left\{
\begin{array}{cc}
a_i \cdot min(x_i), & a_i > 0 \\
a_i \cdot max(x_i), & a_i < 0 \\
\end{array}\right. \\
\end{array}
$$
If $min(e) > c$, then the constraint is inconsistent. Otherwise, for each variable $x_i$ we calculate
the minimum of the expression $e_{x_i}$ obtained from the expression $e$ by removing the monomial $a_i \cdot x_i$, that is $e_{x_i} \equiv a_1 \cdot x_1 + \ldots{} + a_{i-1} \cdot x_{i-1} + a_{i+1} \cdot x_{i+1} + \ldots{} + a_n \cdot x_n$. Such minimum $min(e_{x_i})$ is trivially calculated as $min(e) - min(a_i \cdot x_i)$. Now, for each variable $x_i$ we have that:
\begin{equation}
\label{eq_min_prunings}
x_i \le \left\lfloor\frac{c - min(e_{x_i})}{a_i}\right\rfloor\ \mathrm{if}\ a_i > 0\quad\mathrm{and}\quad
x_i \ge \left\lceil\frac{c - min(e_{x_i})}{a_i}\right\rceil\ \mathrm{if}\ a_i < 0
\end{equation}
The values that do not satisfy the calculated bound should be pruned from the domain of the corresponding variable. These prunings ensure that the constraint is bound consistent. 

A similar analysis can be done for the constraint $e \ge c$, only then the maximums are considered:
$$
max(e) = max(a_1 \cdot x_1) + \ldots{} + max(a_n \cdot x_n)
$$
where 
$$
\begin{array}{ccc}
max(a_i \cdot x_i) & = & \left\{
\begin{array}{cc}
a_i \cdot max(x_i), & a_i > 0 \\
a_i \cdot min(x_i), & a_i < 0 \\
\end{array}\right. \\
\end{array}
$$
If $max(e) < c$, the constraint is inconsistent. Otherwise, the bounds for the variables are calculated as follows:
\begin{equation}
\label{eq_max_prunings}
x_i \ge \left\lceil\frac{c - max(e_{x_i})}{a_i}\right\rceil\ \mathrm{if}\ a_i > 0\quad\mathrm{and}\quad
x_i \le \left\lfloor\frac{c - max(e_{x_i})}{a_i}\right\rfloor\ \mathrm{if}\ a_i < 0
\end{equation}

\begin{algorithm}[!h]
\begin{algorithmic}
\REQUIRE $e = a_1 \cdot x_1 + \ldots{} + a_n \cdot x_n$
\REQUIRE imposed constraint is $e \le c$
\ENSURE $bounds[x_i]$ holds the calculated bound for $x_i$ (upper bound if $a_i > 0$,  lower bound if $a_i < 0$)
\STATE{\textbf{begin}}
\STATE\COMMENT{Let $V(e)$ denote the set of variables appearing in $e$}
\STATE\COMMENT{First, we calculate minimum $min(e)$}
\STATE $min(e) = 0$
\FORALL {$x_i \in V(e)$}
\IF{$a_i > 0$}
\STATE $min(e) = min(e) + a_i \cdot min(x_i)$
\ELSE 
\STATE $min(e) = min(e) + a_i \cdot max(x_i)$
\ENDIF
\ENDFOR
\STATE\COMMENT{Next, we check for inconsistency}
\IF{$min(e) > c$}
\STATE \textbf{return false} \COMMENT{inconsistency is detected}
\ENDIF
\STATE\COMMENT{Finally, we calculate the bounds of the variables in $V(e)$}
\FORALL{$x_i \in V(e)$}
\IF{$a_i > 0$}
\STATE $min(e_{x_i}) = min(e) - a_i \cdot min(x_i)$
\STATE $bounds[x_i] = \left\lfloor \frac{c - min(e_{x_i})}{a_i} \right\rfloor$ \COMMENT{upper bound}
\ELSE 
\STATE $min(e_{x_i}) = min(e) - a_i \cdot max(x_i)$
\STATE $bounds[x_i] = \left\lceil \frac{c - min(e_{x_i})}{a_i} \right\rceil$ \COMMENT{lower bound}
\ENDIF
\ENDFOR
\STATE \textbf{return true} \COMMENT{inconsistency is \emph{not} detected}
\STATE{\textbf{end}}
\end{algorithmic}
\caption{$\mathtt{calculateBoundsStandard}(e, c, bounds)$}
\label{algo:bounds_standard}
\end{algorithm}

\subsection{Improved filtering algorithm} 
\label{subsec:improved}

Assume again the constraint
$e \le c$, where $e \equiv a_1 \cdot x_1 + \ldots{} + a_n \cdot x_n$. The goal is to calculate the \emph{improved minimum} of the expression $e$ (denoted by $min^*(e)$), taking into account the imposed \alldiff constraints on variables in $e$. If that improved minimum is greater than $c$, the constraint is inconsistent. Otherwise, we calculate in the similar fashion the improved minimums 
$min^*(e_{x_i})$ which are then used to calculate the bounds for the variables (in the similar way as in the standard algorithm):
\begin{equation}
\label{improved_min_prunings}
x_i \le \left\lfloor\frac{c - min^*(e_{x_i})}{a_i}\right\rfloor\ \mathrm{if}\ a_i > 0\quad\mathrm{and}\quad
x_i \ge \left\lceil\frac{c - min^*(e_{x_i})}{a_i}\right\rceil\ \mathrm{if}\ a_i < 0
\end{equation}
For efficiency, we would like to avoid calculating each $min^*(e_{x_i})$ from scratch. So there are two main issues to consider here: the first is how to calculate the improved minimum $min^*(e)$, and the second is how to efficiently
calculate the \emph{correction} for $min^*(e_{x_i})$, i.e.~the value $c_i$ such that $min^*(e_{x_i}) = min^*(e) - c_i$.  
The similar situation is with the constraint $e \ge c$, except in that case the improved maximums $max^*(e)$ and $max^*(e_{x_i})$
are calculated, and then the bounds are obtained from the following formulae:
\begin{equation}
\label{improved_max_prunings}
x_i \ge \left\lceil\frac{c - max^*(e_{x_i})}{a_i}\right\rceil\ \mathrm{if}\ a_i > 0\ \mathrm{and}\ 
x_i \le \left\lfloor\frac{c - max^*(e_{x_i})}{a_i}\right\rfloor\ \mathrm{if}\ a_i < 0
\end{equation}

\begin{sloppypar}
We develop the improved algorithm in two stages. First we consider the simple case where all coefficients $a_i$ are of the same sign, and there is the constraint $\texttt{alldifferent}(x_1,\ldots{},x_n)$ in the considered problem (i.e.~all the variables in $e$ must be pairwise distinct).
Then we consider the general case, where there may be more than one \alldiff constraints that (partially) overlap with the variables of $e$, and the coefficients $a_i$ may be both positive or negative.
\end{sloppypar}

\subsubsection{Simple case} 

Assume that $a_i > 0$. The improved minimum $min^*(e)$ is calculated by the procedure $\mathtt{calculateImprovedMinimum}(e, min, M, p)$ (Algorithm \ref{algo:improved_minimum}). The procedure takes the expression $e$ as its input and calculates the improved minimum of $e$
(returned by the output parameter $min$), the \emph{minimizing matching} $M$ and the \emph{next candidate index vector} $p$ (which is later used for calculating corrections). A \emph{matching} $M$ is a sequence of assignments 
$[ x_{i_1} = d_{i_1}, \ldots{}, x_{i_n} = d_{i_n} ]$, where $x_{i_1},\ldots{},x_{i_n}$ is a permutation of $x_1,\ldots{},x_n$, such that $d_{i_j} \ge min(x_{i_j})$ (i.e.~each variable takes a value greater or equal to its minimum) and $d_{i_k} < d_{i_l}$ for $k < l$ (i.e.~the \alldiff constraint is satisfied).\footnote{Notice that a matching $M$ does not have to be a part of any solution of the corresponding CSP, since the maximums of the variables may be violated by $M$.} A matching $M$ is \emph{minimizing} if $\sum_{j = 1}^n a_{i_j} \cdot d_{i_j}$ is as minimal as possible. The procedure constructs such matching and assigns the value $\sum_{j = 1}^n a_{i_j} \cdot d_{i_j}$ to $min$.

\begin{algorithm}[!h]
\begin{algorithmic}
\REQUIRE $e = a_1 \cdot x_1 + \ldots{} + a_n \cdot x_n$
\REQUIRE $a_i > 0$
\ENSURE $M = [ x_{i_1} = d_{i_1}, \ldots{}, x_{i_n} = d_{i_n} ]$ is a minimizing matching
\ENSURE $min = \sum_{j=1}^n a_{i_j} \cdot d_{i_j}$
\ENSURE $p[j]$ is the index in $M$ of the variable which was the next best choice for the value $d_{i_j}$, or $undef$ if 
        there were no other candidates
\STATE{\textbf{begin}}
\STATE \COMMENT{Let $V(e)$ denote the set of variables appearing in $e$}
\STATE $vars = sort(V(e))$ \COMMENT{sort $V(e)$ 	 with respect to the ascending order of minimums}
\STATE $heap.init()$  \COMMENT{$heap$ will contain variables, and the variable with the greatest coefficient will always be on the top of $heap$ (in case of multiple such variables, the one with the smallest index in $e$ will be on the top). Initially, $heap$ is empty}

\STATE $min = 0$
\STATE $M = [\ ]$ \COMMENT{Initially, $M$ is an empty sequence}
\STATE $i = 1$
\STATE $d = min(vars[1]) - 1$
\FOR{$j = 1$ to $n$}
\STATE $d = max(d + 1, min(vars[j]))$  \COMMENT{calculate the next value to be assigned}
\WHILE{$i \le n \wedge min(vars[i]) \le d$}
\STATE $heap.add(vars[i])$  \COMMENT{add candidates for the value $d$ to $heap$}
\STATE $i = i + 1$
\ENDWHILE
\STATE $x = heap.get\_top()$ \COMMENT{remove the top variable from $heap$ (denoted by $x$)}
\STATE $M.push\_back(x = d)$ 
\STATE $min = min + a \cdot d$ \COMMENT{where $a$ is the coefficient for $x$ in $e$}
\IF{$heap$ is not empty}
\STATE $next[j] = heap.view\_top()$ \COMMENT{read and store the next variable from the top of $heap$ without removing it}
\ELSE 
\STATE $next[j] = undef$
\ENDIF
\STATE $index[x] = j$ \COMMENT{store the index of $x$ in $M$}
\ENDFOR
\FOR{$j = 1$ to $n$}
\STATE $p[j] = index[next[j]]$  \COMMENT{with abuse of notation, assume that index of $undef$ is $undef$}
\ENDFOR
\STATE{\textbf{end}}
\end{algorithmic}
\caption{$\mathtt{calculateImprovedMinimum}(e, min, M, p)$}
\label{algo:improved_minimum}
\end{algorithm}

The procedure $\mathtt{calculateImprovedMinimum}()$ works as follows. Let $V(e)$ be the set of variables that appear in $e$. At the beginning of the procedure, we sort the variables from $V(e)$ in the ascending order with respect to their minimums (the vector denoted by $vars$ in Algorithm \ref{algo:improved_minimum}). The main idea is to traverse the variables in $vars$ and assign 
to each variable $x$ the lowest possible yet unassigned value $d \ge min(x)$, favoring the variables with greater coefficients 
whenever possible. For this reason, we maintain the integer variable $d$ which holds the greatest value that is assigned to some 
variable so far (initially it is set to be lower than the lowest minimum, i.e.~it has the value $min(vars[1]) - 1$). In each 
subsequent iteration we calculate the next value to be assigned to some of the remaining variables. It will be the lowest possible value greater than $d$ for which there is at least one \emph{candidate} among the 
remaining variables. The \emph{candidate variables} for some value are those variables whose minimums are lower or 
equal to that value. In case of multiple candidates, we choose the variable with the greatest coefficient, because we want to minimize the sum. If the candidate with the greatest coefficient is not unique (since multiple variables may have equal coefficients in $e$), the variable $x_i$ with the smallest index $i$ in $e$ is chosen.\footnote{Actually, if the candidate with the greatest coefficient is not unique, any of such candidates may be chosen --- the calculated improved minimum would be the same. The rule that chooses the candidate with the smallest index in $e$ is used by the algorithm only to make the execution deterministic.} In order to efficiently find such candidate, we also maintain the $heap$ of candidate variables. Each time we calculate the next value $d$, we add new candidates to the heap (the candidates remained from previous iterations also stay on the heap, since these variables are also 
candidates for the newly calculated value $d$). The variable on the top of the heap is the one
with the greatest coefficient (and with the smallest index).
After we assign the current value $d$ to the variable $x$ removed from the top of the heap, we append the assignment $x = d$ to $M$, add $a \cdot d$ to $min$ (where $a$ is the coefficient for $x$ in $e$) and proceed with the next iteration. At the end of the procedure, the output parameter $min$ has the value $\sum_{j = 1}^n a_{i_j} \cdot d_{i_j}$, where $M = [ x_{i_1} = d_{i_1},\ldots{},x_{i_n} = d_{i_n} ]$, and this value is used as the improved minimum $min^*(e)$. 

After $min^*(e)$ is calculated (and assuming that $min^*(e) \le c$), we want to calculate $min^*(e_{x_i})$ for each $i \in \{1,\ldots{},n\}$, but we want to avoid repeating the algorithm from scratch $n$ times. The idea is to use the obtained improved 
minimum $min^*(e)$ and the minimizing matching $M$ to efficiently reconstruct the result that would be obtained if the 
algorithm was invoked for $e_{x_i}$. Assume that $M = [ x_{i_1} = d_{i_1},\ldots{},x_{i_n} = d_{i_n} ]$, and assume that for some $x_{i_j}$ we want to calculate $min^*(e_{x_{i_j}})$. If we executed the algorithm with $e_{x_{i_j}}$ as input, the obtained 
minimizing matching $M_j$ would coincide with $M$ up to the $j$-th position in the sequence. The first difference
would be at $j$th position, since the variable $x_{i_j}$ did not appear in $e_{x_{i_j}}$, so it would not be on the heap.
In such situation, there are two possible cases:
\begin{itemize}
 \item if the variable $x_{i_j}$ was not the only candidate for the value $d_{i_j}$ during the construction of $M$, in the absence of $x_{i_j}$ the algorithm would choose the next best candidate variable from the heap  and assign the value $d_{i_j}$ to it. Let $x_{i_l}$ be the chosen next best candidate (notice that $l > j$, since $x_{i_l}$ is sequenced after $x_{i_j}$ in $M$).
After the variable $x_{i_l}$ was removed from the heap and the assignment $x_{i_l} = d_{i_j}$ was appended to $M_j$, the remaining 
variables would be arranged in the same way as in the minimizing matching $M_l$, i.e.~the one that would be obtained if the algorithm was 
invoked for $e_{x_{i_l}}$. In other words, the only difference between $M_j$ and $M_l$ is the assignment $x_{i_l} = d_{i_j}$
instead of $x_{i_j} = d_{i_j}$, so we can reduce the problem of finding $M_j$ (and $min^*(e_{x_{i_j}})$) to the problem of 
finding $M_l$ (and $min^*(e_{x_{i_l}})$).
\item if $x_{i_j}$ was the only candidate for the value $d_{i_j}$ during the construction of $M$, in the absence of $x_{i_j}$ the algorithm would not be able to use this value, and it would proceed with the next value $d_{i_{j+1}}$ which would be assigned to the best candidate for that value --- this would be the variable $x_{i_{j+1}}$, as before. In the rest of the algorithm's execution, the remaining variables would be arranged in the same way as in $M$. Therefore, $M_j$ would be the same as $M$ with the assignment $x_{i_j} = d_{i_j}$ removed.
\end{itemize}
In order to be able to reconstruct the described behaviour of the algorithm for $e_{x_{i_j}}$ without invoking it, during the execution of the algorithm for $e$ we remember the second best candidate for each value in the obtained minimizing matching $M$. After the best candidate is removed from the heap and the corresponding assignment is 
appended to $M$, we retrieve the variable on the top of the heap (without removing it) and remember this variable as the second 
best choice for the current value (denoted as $next[j]$ in Algorithm \ref{algo:improved_minimum}). If the heap is empty, we use the
special $undef$ value to denote that there are no alternative candidates. At the end of the algorithm for each value $d_{i_j}$ in
$M$ we calculate $p[j]$ which is the index in $M$ where the variable $next[j]$ is positioned ($p[j] > j$). If there are no other candidates for $d_{i_j}$, then $p[j] = undef$.

Consequently, when calculating the corrections $c[j] = min^*(e) - min^*(e_{x_{i_j}})$, we have two cases. If $p[j] = undef$, the 
minimizing matching $M_j$ would be exactly the same as $M$, with the assignment $x_{i_j} = d_{i_j}$ removed, 
so $c[j] = a_{i_j} \cdot d_{i_j}$. On the other hand, if $p[j] = l > j$, we may assume that $c[l]$ is already calculated (i.e.~we may calculate the corrections in the reversed order). Since the minimizing matching $M_j$ may be reconstructed from the  
minimizing matching $M_l$ by replacing the assignment $x_{i_j} = d_{i_j}$ with the assignment $x_{i_l} = d_{i_j}$, it holds that 
$min^*(e_{x_{i_j}}) = min^*(e_{x_{i_l}}) - a_{i_j} \cdot d_{i_j} + a_{i_l} \cdot d_{i_j} = min^*(e) - c[l] - (a_{i_j} - a_{i_l}) \cdot d_{i_j}$, so $c[j] = (a_{i_j} - a_{i_l}) \cdot d_{i_j} + c[l]$. 

\begin{sloppypar}
The procedure $\mathtt{calculateBoundsImproved}(e, c, bounds)$
(Algorithm \ref{algo:bounds_improved}) has the same parameters and the return value as $\mathtt{calculateBoundsStandard}(e, c, bounds)$, 
but it implements the improved filtering algorithm. It invokes the procedure $\mathtt{calculateImprovedMinimum}(e, min^*(e), M, p)$, checks for 
inconsistency (i.e.~whether $min^*(e) > c$) and then calculates the corrections as previously described. Finally, it calculates the upper bounds for all variables as in the equation \eqref{improved_min_prunings}.
\end{sloppypar}

\begin{algorithm}[!h]
\begin{algorithmic}
\REQUIRE $e = a_1 \cdot x_1 + \ldots{} + a_n \cdot x_n$
\REQUIRE $a_i > 0$
\REQUIRE imposed constraint is $e \le c$
\ENSURE $bounds[x_i]$ holds the calculated upper bound for $x_i$
\STATE{\textbf{begin}}
\STATE\COMMENT{First we calculate the improved minimum $min^*(e)$, the minimizing matching $M$ and the next candidate index vector $p$}
\STATE $\mathtt{calculateImprovedMinimum}(e, min^*(e), M, p)$
\STATE\COMMENT{Now $M$ is the sequence $[ x_{i_1} = d_{i_1}, \ldots{}, x_{i_n} = d_{i_n} ]$}

\STATE\COMMENT{Next, we check for inconsistency}
\IF{$min^*(e) > c$}
\STATE \textbf{return false} \COMMENT{inconsistency is detected}
\ENDIF

\STATE\COMMENT{Finally, we calculate the bounds for the variables in $M$}
\FOR{$j = n$ \textbf{downto} $1$}
\STATE\COMMENT{We calculate the correction $c[j]$}
\IF{$p[j] = undef$}
\STATE $c[j] = a_{i_j} \cdot d_{i_j}$
\ELSE
\STATE $c[j] = d_{i_j} \cdot (a_{i_j} - a_{i_{p[j]}}) + c[p[j]]$
\ENDIF
\STATE $min^*(e_{x_{i_j}}) = min^*(e) - c[j]$
\STATE $bounds[x_{i_j}] = \left\lfloor\frac{c - min^*(e_{x_{i_j}})}{a_{i_j}}\right\rfloor$
\ENDFOR
\STATE \textbf{return true} \COMMENT{inconsistency is \emph{not} detected}
\STATE{\textbf{end}}
\end{algorithmic}
\caption{$\mathtt{calculateBoundsImproved}(e, c, bounds)$}
\label{algo:bounds_improved}
\end{algorithm}

Using the improved minimums. we can often prune more values, as shown in the following example. 

\begin{exa}
\label{ex:improved}
Consider the following CSP: 
$$
\begin{array}{l}
 x_1 \in \{ 1,\ldots{},10\},\ x_2 \in \{ 2,\ldots{},10\} \\
 x_3 \in \{ 1,\ldots{}, 10 \},\ x_4 \in \{ 3, \ldots{}, 10 \} \\
 x_5 \in \{ 3,\ldots{},15 \},\ x_6 \in \{ 9,\ldots{},40\} \\
 \mathtt{alldifferent}(x_1, x_2, x_3, x_4, x_5, x_6) \\
 6x_1 + 8x_2 + 7x_3 + 4x_4 + 2x_5 + x_6 \le 85 
\end{array}
$$
Let $e$ denote the left-hand side expression of the linear constraint in this CSP. If we apply the standard algorithm to calculate bounds (i.e.~we ignore the \alldiff constraint), then $min(e) = 56$, which means that no inconsistency is detected, 
and $min(e_{x_i})$ values and the calculated bounds of the variables are given in the following table:

\begin{center}
\begin{tabular}{|c|c|c|c|c|c|c|}
 \hline
 \textbf{Variable} & $x_1$ & $x_2$ & $x_3$ & $x_4$ & $x_5$ & $x_6$ \\
  \hline
 $min(e_{x_i})$ & 50 & 40 & 49 & 44 & 50 & 47 \\
 \hline
 $x_i \le$ & \textbf{5} & \textbf{5} & \textbf{5} & 10 & 17 & \textbf{38} \\
\hline
\end{tabular}
\end{center}
The bold values denote the calculated bounds that induce prunings. Let us now consider the improved algorithm applied to
the same CSP. First we calculate $min^*(e)$ by invoking the procedure $\mathtt{calculateImprovedMinimum}()$ for $e$:
\begin{itemize}
 \item the first value considered is $d = 1$ for which there are two candidates: $x_1$ and $x_3$. The procedure chooses $x_3$ 
 because its coefficient is greater ($x_1$ stays on the heap as a candidate for the next value). The next best candidate $x_1$ is 
 also remembered for this value. 
 \item the next value considered is $d = 2$ for which there are two candidates: $x_1$ and $x_2$. This time the procedure chooses 
 $x_2$, and $x_1$ will have to wait on the heap for the next value (it is again the next best candidate for $d=2$).
 \item the next value is $d = 3$, and there are three candidates; $x_1$, $x_4$ and $x_5$. The variable $x_1$ is chosen, since its 
 coefficient is the greatest. The next best candidate is $x_4$.
 \item for the next value $d = 4$ there are two candidates on the heap: $x_4$ and $x_5$. The procedure chooses $x_4$, and 
 $x_5$ is remembered as the next best candidate.
 \item the next value is $d = 5$ which is assigned to $x_5$, since it is the only candidate (so the next best candidate is $undef$). 
 \item the next value for which there is at least one candidate is $d = 9$. The only candidate is $x_6$ which is chosen 
 by the procedure (the next best candidate is again $undef$). 
 \end{itemize}
 The obtained matching is $M = [ x_3 = 1, x_2 = 2, x_1 = 3, x_4 = 4, x_5 = 5, x_6 = 9 ]$, the improved minimum is 
 $min^*(e) = 76$ and the next candidate index vector is $p = [ 3, 3, 4, 5, undef, undef ]$ (that is, for the value $1$ the next candidate is $x_1$ which is at the position $3$ in $M$, for the value $2$ the next candidate is $x_1$ which is at the position $3$, for the value $3$ the next candidate is $x_4$ which is at the position $4$ and so on). Since $76 \le 85$, no inconsistency is detected. After the corrections are calculated, we can calculate the improved minimums $min^*(e_{x_i})$ and use these values to calculate the bounds, as shown in the following table:
 
\begin{center}
\begin{tabular}{|c|c|c|c|c|c|c|c|c|}
 \hline
 \textbf{Variable} & $x_1$ & $x_2$ & $x_3$ & $x_4$ & $x_5$ & $x_6$  \\
  \hline
 $j$ (index in $M$) & 3 & 2 & 1 & 4 & 5 & 6 \\
  \hline
  $p[j]$ &   4 & 3 & 3 & 5 & $undef$ & $undef$ \\
  \hline
  $c[j]$ & 24 & 28 & 25 & 18 & 10 &  9 \\ 
  \hline
 $min^*(e_{x_i})$ & 52 & 48 & 51 & 58 & 66 & 67 \\
 \hline
 $x_i \le$ & 5 & \textbf{4} & \textbf{4} & \textbf{6} & \textbf{9} & \textbf{18} \\
\hline
\end{tabular}
\end{center}
The bold values denote the bounds that are stronger compared to those obtained by the standard algorithm. This example
confirms that our algorithm may produce more prunings than the standard algorithm. 
\end{exa}

\paragraph{Complexity.} The complexity of the procedure $\mathtt{calculateImprovedMinimum}()$ is $O(n \log(n))$, since the variables are first sorted once, and then each of the variables is once added and once removed from the heap. The procedure 
$\mathtt{calculateBoundsImproved}()$ invokes the procedure $\mathtt{calculateImprovedMinimum}()$ once, and then calculates the corrections and the bounds in linear time. Thus, its complexity is also $O(n \log(n))$. 

\paragraph{Consistency level.} As we have seen in Example \ref{ex:improved}, our improved algorithm does induce a stronger constraint propagation than the standard algorithm. However, our algorithm does not enforce bound consistency on a conjunction of an \alldiff constraint and a linear constraint. At this point it may be interesting to compare our algorithm to the algorithm developed by Beldiceanu et al.~(\cite{beldiceanu}), which targets a similar problem --- the conjunction $x_1 + \ldots{} + x_n \le c\ \wedge\ \mathtt{alldifferent}(x_1,\ldots{},x_n)$ (notice that it requires that all the coefficients in the sum are equal to $1$). The algorithm establishes bound consistency on such conjunction. The two algorithms have quite similar structures. The main difference is in the ordering used for the candidates on the heap. The algorithm developed by Beldiceanu et al.~(\cite{beldiceanu}) favors the variable with the lowest maximum. The authors show that, in case all coefficient are $1$, this strategy leads to a matching $M$ that minimizes the sum, satisfying both constraints, and respecting \emph{both minimums and maximums} of the variables. If such matching does not exist, the algorithm will report inconsistency. In our algorithm, we allow arbitrary positive coefficients, and the best candidate on the heap is always the one with the greatest coefficient. The matching $M$ will again satisfy both constraints, but only minimums of the variables are respected (maximums may be violated, since they are completely ignored by the algorithm). Because of this relaxation, the calculated improved minimum may be lower than the real minimum, thus some inconsistencies may not be detected, and bound consistency will certainly not be established. In other words, we gave up the bound consistency in order to cover more general case with arbitrary coefficients.

\paragraph{The constraint $e \ge c$.} As said earlier, the case of the constraint $e \ge c$ is similar, except that the improved maximums $max^*(e)$ and $max^*(e_{x_i})$ are considered. To calculate the improved maximum $max^*(e)$, the procedure that is analogous to the procedure $\mathtt{calculateImprovedMinimum}()$ (Algorithm \ref{algo:improved_minimum}) is used, with two main differences. First, the variables are sorted with respect to the descending order of maximums. Second, the value $d$ is initialized to $max(M_1,\ldots{},M_n) + 1$, where $M_i = max(x_i)$, and in each subsequent iteration it takes the greatest possible value lower than its previous value for which there is at least one candidate among the remaining variables (this time, a candidate is a variable $y$ such that $max(y) \ge d$). The corrections for $max^*(e_{x_i})$ are calculated in exactly the same way as in Algorithm \ref{algo:bounds_improved}. The bounds are calculated as stated in the equation \eqref{improved_max_prunings}. 
 
\paragraph{Negative coefficients.} Let us now consider the constraints $e \le c$ and $e \ge c$, where $e \equiv a_1 \cdot x_1 + \ldots{} + a_n \cdot x_n$, and $a_i < 0$. Notice that in this case 
$e \equiv -|e|$, where $|e| \equiv |a_1| \cdot x_1 + \ldots{} + |a_n| \cdot x_n$. For this reason, $min^*(e) = -max^*(|e|)$, and
similarly, $max^*(e) = -min^*(|e|)$, so the problem of finding improved minimums/maximums is easily reduced to the previous case with the positive coefficients. To calculate corrections for $min^*(e_{x_i})$ and $max^*(e_{x_i})$, we could calculate 
the corrections for $min^*(|e_{x_i}|)$ and $max^*(|e_{x_i}|)$ as in Algorithm \ref{algo:bounds_improved}, only with the sign changed. 
 
\subsubsection{General case}
\label{subsubsec:general}

Assume now the general case, where the variables may have both positive and negative coefficients in $e$, and there are multiple \alldiff constraints that partially overlap with the variables in $V(e)$. We reduce this general case to the previous simple case by partitioning the set $V(e)$ into disjoint subsets to which the previous algorithms may be applied.
The partitioning is done by the procedure $\mathtt{findPartitions}(e, csts, pts)$ (Algorithm \ref{algo:partition}). 
This procedure takes the expression $e$ and the set of all \alldiff constraints in the considered problem (denoted by $csts$) as inputs. Its output parameter is $pts$ which will hold the result. The first step is to partition the 
set $V(e)$ as $V^+ \cup V^-$, where $V^+$ contains all the variables from $V(e)$ with positive coefficients, and $V^-$ 
contains all the variables from $V(e)$ with negative coefficients. Each set $V \in \{ V^+, V^- \}$ is further partitioned into \emph{disjoint} $ad$-\emph{partitions}. We say that the set of variables $V' \subseteq V$ is an \alldiff \emph{partition} (or $ad$-\emph{partition}) in $V$ if there 
is an \alldiff constraint in the considered CSP problem that includes all variables from $V'$. Larger $ad$-partitions are preferred, so we first look for the largest $ad$-partition in $V$ by considering all intersections of $V$ with the relevant \alldiff constraints and choosing the one with the largest cardinality. When such $ad$-partition is identified, the variables that make the partition are removed from $V$ and the remaining $ad$-partitions are then searched for in the same fashion among the rest of the variables in $V$. It is very important to notice that the obtained partitions depend only on the \alldiff constraints in the problem and the signs of the coefficients in $e$. Since these are not changed during solving, each invocation of the procedure 
$\mathtt{findPartitions}()$ would yield the same result. This means that the procedure may be called \emph{only once} at the beginning of the solving process, and its result may be stored and used later when needed. 

\begin{algorithm}[!h]
\begin{algorithmic}
\REQUIRE $e = a_1 \cdot x_1 + \ldots{} + a_n \cdot x_n$
\REQUIRE $csts$ in the set of the \alldiff constraints in the considered problem
\ENSURE $pts$ will be the set of disjoint subsets of $V(e)$ such that each $x_i \in V(e)$ belongs to some $S \in pts$
\ENSURE for each $S \in pts$, all variables in $S$ have coefficients of the same sign in $e$
\ENSURE for each $S \in pts$ such that $|S| > 1$, there is an \alldiff constraint in $csts$ that covers all the variables in $S$
\STATE{\textbf{begin}}
\STATE \COMMENT{Let $V(e) = V^+ \cup V^-$, where $V^+$ is the set of variables with positive coefficients in $e$, and 
$V^-$ is the set of variables with negative coefficients in $e$}
\STATE $pts = \varnothing$
\STATE \COMMENT{The partitioning is done separately for $V^+$ and $V^-$}
\FORALL{$V \in \{ V^+, V^- \}$}
\STATE $V_{curr} = V$
\WHILE{$V_{curr} \ne \varnothing$}
\STATE $S_{max} = \varnothing$
\FORALL{$ad \in csts$}
\STATE $S_{curr} = V(ad) \cap V_{curr}$ \COMMENT{$V(ad)$ denotes the set of variables of the constraint $ad$}
\IF{$|S_{curr}| > |S_{max}|$}
\STATE $S_{max} = S_{curr}$
\ENDIF
\ENDFOR
\IF{$|S_{max}| \ge 2$}
\STATE \COMMENT{The case when the largest $ad$-partition has at least $2$ variables}
\STATE $pts = pts \cup \{ S_{max} \}$
\STATE $V_{curr} = V_{curr} \setminus S_{max}$
\ELSE 
\STATE \COMMENT{The case when all the remaining variables are singleton partitions}
\FORALL{$x \in V_{curr}$}
\STATE $S = \{ x \}$
\STATE $pts = pts \cup \{ S \}$
\ENDFOR
\STATE $V_{curr} = \varnothing$
\ENDIF
\ENDWHILE
\ENDFOR
\STATE{\textbf{end}}
\end{algorithmic}
\caption{$\mathtt{findPartitions}(e, csts, pts)$}
\label{algo:partition}
\end{algorithm}

\begin{sloppypar}
The procedure $\mathtt{calculateBoundsImprovedGen}(e, c, bounds)$ (Algorithm \ref{algo:bounds_improved_gen}) has the same 
parameters and the return value as the procedure $\mathtt{calculateBoundsImproved}()$ (Algorithm \ref{algo:bounds_improved}), but it is suitable for the general case. It depends on the partitioning for $e \equiv e^1 + \ldots{} + e^s$ obtained by the procedure $\mathtt{findPartitions}()$ (Algorithm \ref{algo:partition}). Since the variables in each partition $e^k$ have the coefficients of the same sign, and are also covered by some of the \alldiff constraints in the considered CSP, the procedure $\mathtt{calculateImprovedMinimum}()$ (Algorithm \ref{algo:improved_minimum}) may be used to calculate the improved minimums
$min^*(e^k)$. The improved minimum $min^*(e)$ calculated by the procedure $\mathtt{calculateBoundsImprovedGen}()$ is 
the sum of the improved minimums for $e^k$, that is: $min^*(e) = \sum_{k=1}^s min^*(e^k)$. If $min^*(e) > c$, the procedure $\mathtt{calculateBoundsImprovedGen}()$ reports 
an inconsistency. Otherwise, the procedure calculates the corrections and the bounds in an analogous way as in Algorithm \ref{algo:bounds_improved}. There are two important differences. First, we use the value $b^k = c - (min^*(e) - min^*(e^k))$  as the upper bound for the expression $e^k$ instead of $c$ when calculating the bounds for the variables of $e^k$. Another difference is that we must distinguish two cases, depending on the sign of the coefficient: for positive coefficients the calculated bound is the upper bound, and for negative it is the lower bound (because the orientation of the inequality is reversed when dividing with the negative coefficient). 
\end{sloppypar}

\begin{algorithm}[!h]
\begin{algorithmic}
\REQUIRE $e = a_1 \cdot x_1 + \ldots{} + a_n \cdot x_n$
\REQUIRE imposed constraint is $e \le c$
\ENSURE $bounds[x_i]$ holds the calculated bound for $x_i$ (upper bound for $a_i > 0$, lower bound for $a_i < 0$)
\STATE{\textbf{begin}}
\STATE \COMMENT{Let $e \equiv e^1 + \ldots{} + e^s$, where $e^k$ are the partitions of $e$ obtained by Algorithm \ref{algo:partition}}
\STATE \COMMENT{First, we calculate the improved minimum for $e$}
\STATE $min^*(e) = 0$
\FOR{$k = 1$ \textbf{to} $s$}
\STATE $\mathtt{calculateImprovedMinimum}(e^k, min^*(e^k), M^k, p^k)$
\STATE $min^*(e) = min^*(e) + min^*(e^k)$
\ENDFOR

\STATE\COMMENT{Next, we check for inconsistency}
\IF{$min^*(e) > c$}
\STATE \textbf{return false} \COMMENT{inconsistency is detected}
\ENDIF

\STATE\COMMENT{For each partition $e^k$ we calculate the corrections and the bounds in a similar way as in Algorithm \ref{algo:bounds_improved}}
\FOR{$k = 1$ \textbf{to} $s$}
\STATE $b^k = c - (min^*(e) - min^*(e^k))$ \COMMENT{$b^k$ is the calculated upper bound for $e^k$}
\STATE\COMMENT{Let $M^k = [x_{i_{k,1}} = d_{i_{k,1}}, \ldots{}, x_{i_{k,n_k}} = d_{i_{k,n_k}}]$}
\FOR{$j = n_k$ \textbf{downto} $1$}
\STATE\COMMENT{We calculate the correction $c^k[j]$}
\IF{$p^k[j] = undef$}
\STATE $c^k[j] = a_{i_{k,j}} \cdot d_{i_{k,j}}$
\ELSE
\STATE $c^k[j] = d_{i_{k,j}} \cdot (a_{i_{k,j}} - a_{i_{k,p^k[j]}}) + c^k[p^k[j]]$
\ENDIF
\STATE $min^*(e^k_{x_{i_{k,j}}}) = min^*(e^k) - c^k[j]$
\IF{$a_{i_{k,j}} > 0$}
\STATE $bounds[x_{i_{k,j}}] = \left\lfloor\frac{b^k - min^*(e^k_{x_{i_{k,j}}})}{a_{i_{k,j}}}\right\rfloor$ \COMMENT{upper bound}
\ELSE
\STATE $bounds[x_{i_{k,j}}] = \left\lceil\frac{b^k - min^*(e^k_{x_{i_{k,j}}})}{a_{i_{k,j}}}\right\rceil$ \COMMENT{lower bound}
\ENDIF
\ENDFOR
\ENDFOR
\STATE \textbf{return true} \COMMENT{inconsistency is \emph{not} detected}
\STATE{\textbf{end}}
\end{algorithmic}
\caption{$\mathtt{calculateBoundsImprovedGen}(e, c, bounds)$}
\label{algo:bounds_improved_gen}
\end{algorithm}

\begin{sloppypar}
\paragraph{Complexity.} Let $n_k = |V(e^k)|$, and let $n = |V(e)| = n_1 + \ldots{} + n_s$. The procedure $\mathtt{calculateBoundsImprovedGen}()$ (Algorithm \ref{algo:bounds_improved_gen}) invokes the procedure $\mathtt{calculateImprovedMinimum}()$ (Algorithm \ref{algo:improved_minimum}) once for each partition $e^k$, which takes 
$\sum_{k=1}^s O(n_k \log(n_k)) = O(n \log(n))$ time. Since the corrections and the bounds are calculated in linear time (with respect to $n$), the total complexity is $O(n \log(n))$. 
\end{sloppypar}

\subsubsection{Algorithm correctness}

In the following text we prove the correctness of the above algorithms. We will use the following notation. For a linear expression $e \equiv a_1\cdot x_1 + \ldots{} + a_n \cdot x_n$ and an $n$-tuple $\mathbf{d} = (d_1,\ldots{},d_n)$ we denote $e[\mathbf{d}] = \sum_{i=1}^n a_i \cdot d_i$. For a matching $M = [ x_{i_1} = d_{i_1}, \ldots{}, x_{i_n} = d_{i_n} ]$ over the set of variables $V(e)$, let $\mathbf{d}(M)$ be the $n$-tuple that corresponds to $M$. We also denote $e(M) = e[\mathbf{d}(M)] = \sum_{j = 1}^n a_{i_j} \cdot d_{i_j}$. Recall that the matchings are always ordered with respect to the assigned values, i.e.~$k < l \Rightarrow d_{i_k} < d_{i_l}$. This implies that for each $n$-tuple $\mathbf{d} = (d_1,\ldots{},d_n)$ where $i \ne j \Rightarrow d_i \ne d_j$, there exists exactly one matching $M = M(\mathbf{d})$ such that $x_i = d_i$ belongs to $M$. 

The correctness of the procedure $\mathtt{calculateImprovedMinimum}()$ (Algorithm \ref{algo:improved_minimum}) follows from Theorem \ref{thm:improved_min} which states 
that the improved minimum obtained by the procedure is indeed the minimal value that the considered expression may take, when all its variables take pairwise distinct values greater or equal to their minimums. 

\begin{thm}
 \label{thm:improved_min}
 Let $e \equiv a_1 \cdot x_1 + \ldots{} + a_n \cdot x_n$, where $a_i > 0$, and let $m_i = min(x_i)$ be the minimum of the variable $x_i$.  Let $M = [ x_{i_1} = d_{i_1}, \ldots{}, x_{i_n} = d_{i_n} ]$ be the matching obtained by invoking the procedure $\mathtt{calculateImprovedMinimum}()$ for the expression $e$, and let $min = e(M) = \sum_{j=1}^n a_{i_j} \cdot d_{i_j}$ be the obtained improved minimum. Then for any $n$-tuple $\mathbf{d} = (d_1,\ldots{}, d_n)$ such that $d_i \ge m_i$ and $d_i \ne d_j$ for $i \ne j$, it holds that $e[\mathbf{d}] \ge min$. 
\end{thm}

\proof
 Let $T = \{ (d_1,\ldots{},d_n)\ |\ d_i \ge m_i,\ i \ne j \Rightarrow d_i \ne d_j \}$ be the set of $n$-tuples satisfying the 
 conditions of the theorem, and let $E(T) = \{ e[\mathbf{d}]\ |\ \mathbf{d} \in T \}$.  The set $T$ is non-empty (for instance, $(h, h+1,h+2,\ldots{},h + n - 1) \in T$, where $h = max(m_1,\ldots{},m_n)$), and the set $E(T)$ is bounded from below (for instance, by $\sum_{i = 1}^n a_i \cdot m_i$). Thus, the minimum $min(E(T))$ is finite, and there is at least one $n$-tuple $\mathbf{d} \in T$ for which $e[\mathbf{d}] = min(E(T))$. We want to prove that the $n$-tuple $\mathbf{d}(M)$ that corresponds to the matching $M$ obtained by the algorithm is one such $n$-tuple, i.e.~that $e(M) = min(E(T))$. First, notice that $\mathbf{d}(M) \in T$ for any matching $M$ (this follows from the definitions of a matching and the set $T$). The state of the algorithm's execution before each iteration of the main loop may be described by a triplet $(V^c, M^c, d^c)$, where $V^c$ is the set of variables to which values have not been assigned yet (initially $\{ x_1,\ldots{}, x_n \}$), $M^c$ is the current partial matching (initially the empty sequence $[\ ]$), and $d^c$ is the maximal value that has been assigned to some of the variables
 so far (initially $min(m_1,\ldots{},m_n) - 1$).
 We will prove the following property: for any state $(V^c, M^c, d^c)$ obtained from the initial state by executing the algorithm, the partial matching $M^c$ may be extended to a full matching $M'$ by assigning the values greater than $d^c$ to the variables from $V^c$, such that $e(M') = min(E(T))$. The property will be proved by induction on $k = |M^c|$. For $k = 0$ the property trivially holds, since the empty sequence may certainly be extended to some minimizing matching using the values greater than $d^c = min(m_1,\ldots{},m_n) - 1$ (because such matching exists). Let us assume that this property holds for the state $(V^c, M^c, d^c)$, where $|M^c| = k$, and prove that the property also holds for $(V^c \setminus \{ \overline{x} \}, [M^c, \overline{x} = \overline{d}], \overline{d})$, where $\overline{x} = \overline{d}$ is the assignment that our algorithm chooses at that point of its execution. Recall that the value $\overline{d}$ chosen by the algorithm is the lowest possible value greater than $d^c$ such that there is at least one candidate among the variables in $V^c$ (i.e.~some $y \in V^c$ such that $min(y) \le \overline{d}$), and that the variable $\overline{x}$ chosen by the algorithm is one of the candidates for $\overline{d}$ with the greatest coefficient (if such candidate is not unique, the algorithm chooses the one with the smallest index in $e$). If $M'$ is any full minimizing matching that extends $M^c$ using the values greater than $d^c$, then the following properties hold:
 \begin{itemize}  
   \item the value $\overline{d}$ must be used in $M'$. Indeed, since $\overline{d}$ has at least one candidate $y \in V^c$, 
   not using $\overline{d}$ in $M'$ means that $y = d' \in M'$, where $d' > \overline{d}$. This is because
   all values assigned to the variables from $V^c$ are greater than $d^c$, and $\overline{d}$ is the lowest possible value 
   greater than $d^c$ with at least one candidate. Because of this, the assignment $y = d' \in M'$ could be substituted 
   by $y = \overline{d}$, obtaining another full matching $M''$ such that $e(M'') < e(M')$.
   This is in contradiction with the fact that $M'$ is a minimizing matching. 
   \item if $y = \overline{d} \in M'$, the variable $y$ must be a candidate for $\overline{d}$ with the greatest possible coefficient. Indeed, if the coefficient of $y$ is $a$, and there is another candidate $y' \in V^c$ with a greater coefficient $a'$, then choosing $y$ instead of $y'$ for the value $\overline{d}$ would again result in a non-minimizing matching $M'$: since $y' = d' \in M'$, where $d' > \overline{d}$, and $y = \overline{d} \in M'$, exchanging the values of $y$ and $y'$ would result in another matching $M''$ such that $e(M'') < e(M')$.
   \item if the candidate for $\overline{d}$ with the greatest coefficient is not unique, then for any such candidate $z$ the assignment $z = \overline{d}$ belongs to some minimizing matching $M''$ that extends $M^c$. Indeed, if $y = \overline{d} \in M'$ and $z$ is another candidate for $\overline{d}$ with the same coefficient as $y$ (that has a value $d' > \overline{d}$ in $M'$), exchanging the values for $y$ and $z$ would result in another matching $M''$ 
   such that the value of the expression $e$ is not changed, i.e.~$e(M'') = e(M') = min(E(T))$.
 \end{itemize}
 From the proven properties  it follows that the assignment $\overline{x} = \overline{d}$ chosen by our algorithm also belongs to some minimizing matching $M'$ that extends $M^c$ by assigning the values greater than $d^c$ to the variables from $V^c$. Since the lowest value greater than $d^c$ with at least one candidate is $\overline{d}$, and since $\overline{x} = \overline{d} \in M'$, it follows that the values assigned to the variables from $V^c \setminus \{ \overline{x} \}$
 must be greater than $\overline{d}$ in the matching $M'$. This proves that the partial matching $[M^c, \overline{x} = \overline{d}]$ can also be extended to the same minimizing matching $M'$ using the values greater than $\overline{d}$. This 
 way we have proven that the stated property holds for the state $(V^c \setminus \{ \overline{x} \}, [M^c, \overline{x} = \overline{d}], \overline{d})$ which is the next state obtained by our algorithm, and $|[M^c, \overline{x} = \overline{d}]| = k + 1$. It now follows by induction that the stated property holds for any state $(V^c, M^c, d^c)$ obtained by the execution of the algorithm, including the final state 
 $(\{\ \}, M, d_{i_n})$, where $M = [ x_{i_1} = d_{i_1}, \ldots{}, x_{i_n} = d_{i_n} ]$. Therefore, $min = e(M) = min(E(T))$, which proves the theorem.
\qed

Let $subst(M, A, B)$ denote the result obtained when a subsequence $A$ is substituted by a (possibly empty) subsequence $B$ in a matching $M$. The following Theorem \ref{thm:corrections} states that the corrections in the procedure $\mathtt{calculateBoundsImproved}()$ (Algorithm \ref{algo:bounds_improved}) are calculated correctly. 

\begin{sloppypar}
\begin{thm}
 \label{thm:corrections}
 Let $e \equiv a_1 \cdot x_1 + \ldots{} + a_n \cdot x_n$, where $a_i > 0$. Let $M = [x_{i_1} = d_{i_1},\ldots{},x_{i_n} = d_{i_n}]$ be the minimizing matching obtained by 
 invoking the procedure $\mathtt{calculateImprovedMinimum}()$ for $e$, and let $min^*(e) = e(M)$ be the calculated improved 
 minimum. Let $M_j$ be the minimizing matching that would be obtained if the procedure was invoked for $e_{x_{i_j}}$ and let $min^*(e_{x_{i_j}}) = e(M_j)$ be the corresponding improved minimum. Finally, let $c_j = min^*(e) - min^*(e_{x_{i_j}})$. If 
 $x_{i_j}$ was the only candidate for the value $d_{i_j}$ when the procedure was invoked for $e$, then $M_j = subst(M, [ x_{i_j} = d_{i_j} ] \rightarrow [\ ])$. Otherwise, if the second best candidate was $x_{i_l}$ ($l > j$), then $M_j = subst(M_l, [ x_{i_j} = d_{i_j} ] \rightarrow [ x_{i_l} = d_{i_j} ])$. Consequently, the corresponding correction is $c_j = a_{i_j} \cdot d_{i_j}$, or $c_j = (a_{i_j} - a_{i_l})\cdot d_{i_j} + c_l$, respectively. 
\end{thm}
\end{sloppypar}

\proof
 First, let $\mathbf{S} = (V, M, d)$ be any state of the execution of the algorithm, with the same meaning as in the proof of Theorem \ref{thm:improved_min}. The \emph{transition relation} between the states will be denoted by $\rightarrow$: if $\mathbf{S}'$ is the next state for $\mathbf{S}$, this will be written as $\mathbf{S} \rightarrow \mathbf{S}'$. The transitive closure of this relation will be denoted by $\rightarrow^*$. It is clear that the relation $\rightarrow$ is deterministic, that is, for any state $\mathbf{S}$ there is a unique state $\mathbf{S}'$ such that $\mathbf{S} \rightarrow \mathbf{S}'$. This is, therefore, also true for the relation $\rightarrow^*$. This means that the state fully determines the further execution of the algorithm --- for two identical states the rest of the execution would be identical. But even for two different states the further execution 
 of the algorithm may partially or fully coincide. We say that two states $\mathbf{S}' = (V', M', d')$ and $\mathbf{S}'' = (V'', M'', d'')$ are \emph{weakly similar} if the assignment that is chosen next by the algorithm is the same for both states, i.e.~there is some assignment $\overline{x} = \overline{d}$ such that $(V', M', d') \rightarrow (V' \setminus \{ \overline{x} \}, [M', \overline{x} = \overline{d}], \overline{d})$, and $(V'', M'', d'') \rightarrow (V'' \setminus \{ \overline{x} \}, [M'', \overline{x} = \overline{d}], \overline{d})$. On the other hand, we say that $\mathbf{S}'$ and $\mathbf{S''}$ are \emph{similar} if \emph{all} the remaining assignments would be chosen in exactly the same 
 way for both states, i.e.~there is some sequence of assignments $S$ such that $(V', M', d') \rightarrow^* (\{\ \}, [M', S], d_f)$, and $(V'', M'', d'') \rightarrow^* (\{\ \}, [M'', S], d_f)$, where $d_f$ is the last value assigned in both matchings. First, notice that the following two properties hold:
 \begin{description}
  \item[Property 1] if for two states $(V', M', d')$ and $(V'', M'', d'')$ we have that $d' = d''$ and $V'' = V' \setminus \{ y \}$, where $y \in V'$ is 
  not the variable chosen by the algorithm for the next assignment in the state $(V',M',d')$, then these two states are
  weakly similar. Indeed, since in the state $(V',M',d')$ the algorithm chooses some assignment $\overline{x} = \overline{d}$, where $\overline{x} \ne y$ and $\overline{d} > d'$, the same assignment will be chosen in the state
  $(V'', M'', d'')$, since $\overline{x} \in V''$, and $d'' = d'$. 
  \item[Property 2] if for two states $(V', M', d')$ and $(V'', M'', d'')$ we have that  
 $V' = V''$ and $d' = d''$, then these two states are similar. Indeed, the existing assignments in $M'$ and $M''$ do not affect the future assignments, which depend only on the remaining variables (and $V' = V''$), and the last assigned value  (and $d' = d''$). 
 \end{description}
Now let $M = [x_{i_1} = d_{i_1},\ldots{},x_{i_n} = d_{i_n}]$ be the minimizing matching obtained by the algorithm for the expression $e$, and let $M_j$ be the minimizing matching obtained by the algorithm for the expression $e_{x_{i_j}}$.
Recall that $e_{x_{i_j}}$ is obtained by removing the monomial $a_{i_j} \cdot x_{i_j}$ from $e$, i.e.~$V(e_{x_{i_j}}) = 
V(e) \setminus \{ x_{i_j} \}$. Let $\mathbf{S}^k$ be the state after $k$ iterations of the algorithm invoked for the expression $e$, and let $\mathbf{S}_j^k$ be the state after $k$ iterations of 
the algorithm invoked for the expression $e_{x_{i_j}}$. Since 
the variable $x_{i_j}$ is not the best candidate until the $j$th position in $M$, according to the first property,
the states $\mathbf{S}^k$ and $\mathbf{S}_j^k$ are weakly similar for $k < j - 1$. In other words, the 
assignments chosen by the algorithm before the $j$th position will be identical in $M$ and $M_j$.   This means that $P = [x_{i_1} = d_{i_1}, \ldots{}, x_{i_{j-1}} = d_{i_{j-1}}]$ is a common prefix for the matchings $M$ and $M_j$. 

Let us now consider the behaviour of the algorithm at $j$th position in more details. Let $V^k = V(e) \setminus \{ x_{i_1}, \ldots{}, x_{i_k} \}$ for all $k \le n$. When the algorithm is invoked for $e$, we have the following state transitions: $\mathbf{S}^{j-1} \rightarrow \mathbf{S}^j \rightarrow \mathbf{S}^{j+1}$, 
where $\mathbf{S}^{j-1} = (V^{j-1}, P, d_{i_{j-1}})$,  $\mathbf{S}^j = (V^j, [P, x_{i_j} = d_{i_j}], d_{i_j})$, 
and $\mathbf{S}^{j+1} = (V^{j+1}, [P,x_{i_j} = d_{i_j}, x_{i_{j+1}} = d_{i_{j+1}}], d_{i_{j+1}})$. On the other hand, if the algorithm 
is invoked for $e_{x_{i_j}}$, when the state $\mathbf{S}_j^{j-1} = (V^{j-1} \setminus \{ x_{i_j} \}, P, d_{i_{j-1}})$ is reached, there are two possible cases. If $x_{i_j}$ is the only candidate for $d_{i_j}$ in the 
state $\mathbf{S}^{j-1}$, then there are no candidates for $d_{i_j}$ in the state $\mathbf{S}_j^{j-1}$. 
Thus, the algorithm proceeds with the first greater value for which there are candidates in $V^{j-1} \setminus \{ x_{i_j} \}$, and this is $d_{i_{j+1}}$. The best candidate for such value will be the same as in the state $\mathbf{S}^j$, because the sets of the remaining variables are the same in both states ($V^{j-1} \setminus \{ x_{i_j} \} = V^j$).  Therefore, the next state will be $\mathbf{S}_j^j = (V^{j-1}  \setminus \{ x_{i_j} \} \setminus \{ x_{i_{j+1}} \}, [P, x_{i_{j+1}} = d_{i_{j+1}}], d_{i_{j+1}})$. Since, according to the second property, the states $\mathbf{S}^{j+1}$ and $\mathbf{S}_j^j$ are similar (because $V^{j-1}  \setminus \{ x_{i_j} \} \setminus \{ x_{i_{j+1}} \} = V^{j+1}$), the final matchings $M$ and $M_j$ will coincide at the remaining positions, that is, $\mathbf{S}^{j+1} \rightarrow^* (\{\ \}, [P, x_{i_j} = d_{i_j}, x_{i_{j+1}} = d_{i_{j+1}}, S], d_{i_n})$ and $\mathbf{S}_j^j \rightarrow^* (\{\ \}, [P, x_{i_{j+1}} = d_{i_{j+1}}, S], d_{i_n})$, where $S = [x_{i_{j+2}} = d_{i_{j+2}}, \ldots{}, x_{i_n} = d_{i_n}]$ is a common suffix of $M$ and $M_j$. Thus, $M_j = subst(M, [x_{i_j} = d_{i_j}] \rightarrow [\ ])$. The corresponding correction is $c_j = min^*(e) - min^*(e_{x_{i_j}}) = e(M) - e_{x_{i_j}}(M_j) = a_{i_j} \cdot d_{i_j}$. On the other 
hand, if there exists the second best candidate $x_{i_l}$ for the value $d_{i_j}$ ($l > j$) in the state $\mathbf{S}^{j-1}$,
this will be the best candidate for $d_{i_j}$ in the state $\mathbf{S}_j^{j-1}$. Thus, the algorithm will choose the 
assignment $x_{i_l} = d_{i_j}$, i.e.~$\mathbf{S}_j^j =  (V^{j-1}  \setminus \{ x_{i_j} \} \setminus \{ x_{i_l} \}, [P, x_{i_l} = d_{i_j}], d_{i_j})$. In order to determine the further state transitions, let us now consider the behaviour of the 
algorithm when invoked for $e_{x_{i_l}}$. As previously stated, the matchings $M_l$ and $M$ will coincide until the 
$l$th position, and since $l > j$, it means that $M_l$ will also have $P$ as its prefix. Therefore, $\mathbf{S}_l^{j-1} = (V^{j-1} \setminus \{ x_{i_l} \}, P, d_{i_{j-1}})$. This state is weakly similar with
$\mathbf{S}^{j-1}$ (according to the first property), so we have the next state 
$\mathbf{S}_l^j = (V^{j-1}  \setminus \{ x_{i_l} \} \setminus \{ x_{i_j} \}, [P, x_{i_j} = d_{i_j}], d_{i_j})$. 
Now states $\mathbf{S}_l^j$ and $\mathbf{S}_j^j$ are obviously similar (according to the second property), so we have $\mathbf{S}_j^j \rightarrow^* (\{\ \}, [P, x_{i_l} = d_{i_j}, S'], d')$ and 
$\mathbf{S}_l^j \rightarrow^* (\{\ \}, [P, x_{i_j} = d_{i_j}, S'], d')$, where $d'$ is the greatest assigned value and $S'$ is a common suffix. This implies that $M_j = subst(M_l, [x_{i_j} = d_{i_j}] \rightarrow [x_{i_l} = d_{i_j}])$. The corresponding correction is $c_j = min^*(e) - min^*(e_{x_{i_j}})
 = e(M) - e_{x_{i_j}}(M_j) = e(M) - (e_{x_{i_l}}(M_l) - a_{i_j} \cdot d_{i_j} + a_{i_l} \cdot d_{i_j}) = 
 c_l + (a_{i_j} - a_{i_l}) \cdot d_{i_j}$, as stated. 
\qed

\noindent{}Now, the following theorem proves the correctness of Algorithm \ref{algo:bounds_improved} (the simple case).

\begin{sloppypar}
\begin{thm}
\label{thm:main}
 Let $P$ be a CSP that contains the constraints $\mathtt{alldifferent}(x_1,\ldots{},x_n)$, and $e \le c$, where $e \equiv a_1 \cdot x_1 + \ldots{} + a_n \cdot x_n$ and $a_i > 0$. If the procedure $\mathtt{calculateBoundsImproved}()$ applied to the expression $e$ returns \emph{false}, then the problem $P$ has no solutions. Otherwise, let $P'$ be the CSP obtained from $P$ by pruning the values outside of the calculated bounds for the variables in $V(e)$. Then $P$ and $P'$ are equivalent.
\end{thm}
\end{sloppypar}
\proof
Let $\mathbf{d} = (d_1,\ldots{},d_n)$ be any $n$-tuple from $\D = (D_{x_1},\ldots{},D_{x_n})$, and let $min^*(e)$ be the improved minimum obtained by the procedure $\mathtt{calculateImprovedMinimum}()$ for the expression $e$. First, notice that if $\mathbf{d}$ satisfies the constraint $\mathtt{alldifferent}(x_1,\ldots{},x_n)$, then $min^*(e) \le  e[\mathbf{d}]$.
Indeed, since $\mathbf{d} \in \D$, it follows that $d_i \ge m_i$, where $m_i = min(x_i)$. Furthermore, since $\mathbf{d}$ satisfies the constraint $\mathtt{alldifferent}(x_1,\ldots{},x_n)$, it follows that $i \ne j \Rightarrow d_i \ne d_j$. This means that the conditions of Theorem \ref{thm:improved_min} are satisfied, so it must be $e[\mathbf{d}] \ge min^*(e)$. 
Similarly, it holds that $e_{x_i}[\mathbf{d}] \ge min^*(e_{x_i})$.

\begin{sloppypar}
If the procedure $\mathtt{calculateBoundsImproved}()$ returns \emph{false} when applied to $e$, this means that $min^*(e) > c$. Thus, for any $\mathbf{d} \in \D$ satisfying the constraint $\mathtt{alldifferent}(x_1,\ldots{},x_n)$ it will be 
$e[\mathbf{d}] \ge min^*(e) > c$, which means that no $n$-tuple $\mathbf{d} \in \D$ satisfies both the \alldiff constraint and the constraint $e \le c$. In other words, $P$ has no solutions.
\end{sloppypar}

Now, to prove the second part of the theorem, we must prove that any value pruned by the improved algorithm does not belong to any solution of $P$. Assume that for some variable $x_i$ a value $d_i$ is pruned from its domain by our improved algorithm. According
to the equation \eqref{improved_min_prunings}, it means that $d_i > \left\lfloor\frac{c - min^*(e_{x_i})}{a_i}\right\rfloor$, or equivalently, $min^*(e_{x_i}) + a_i \cdot d_i > c$. Therefore, for any $n$-tuple $\mathbf{d} \in \D$ that contains $d_i$ as the 
value for $x_i$ and that satisfies the constraint $\mathtt{alldifferent}(x_1,\ldots{},x_n)$, it holds that  $e[\mathbf{d}] = e_{x_i}[\mathbf{d}] + a_i \cdot d_i \ge  min^*(e_{x_i}) + a_i \cdot d_i >  c$ (since $e_{x_i}[\mathbf{d}] \ge min^*(e_{x_i})$), so the constraint
$e \le c$ is not satisfied and $\mathbf{d}$ is not a solution of $P$. This means that there are no solutions of $P$ that
contain $d_i$ as the value for $x_i$ and that satisfy both the constraint $\mathtt{alldifferent}(x_1,\ldots{},x_n)$ and the constraint $e \le c$. This proves the theorem, since all the solutions of $P$ are also the solutions of $P'$. 
\qed

 \noindent{}Finally, we prove the correctness of Algorithm \ref{algo:bounds_improved_gen} (the general case).
 
\begin{thm}
\label{thm:main_gen}
 Let $P$ be a CSP that contains the constraint $e \le c$, where $e \equiv a_1 \cdot x_1 + \ldots{} + a_n \cdot x_n$ ($a_i$ may be both positive or negative) and one or more \alldiff constraints that may overlap with $V(e)$. If the procedure 
 $\mathtt{calculateBoundsImprovedGen}()$ applied to the expression $e$ returns \emph{false}, then the problem $P$ has no solutions. Otherwise, let $P'$ be the CSP obtained from $P$ by pruning the values outside of the calculated bounds for the variables in $V(e)$. Then $P$ and $P'$ are equivalent.
\end{thm}
\proof
Let $e \equiv e^1 + \ldots{} + e^s$ be the partitioning of the expression $e$ obtained by the procedure $\mathtt{findPartitions}()$ (Algorithm \ref{algo:partition}). Since for each partition $e^k$ there is a covering \alldiff constraint in $P$, and since all its variables have the coefficients of the same sign, from Theorem \ref{thm:improved_min} it easily follows that for any $\mathbf{d} \in \D = (D_{x_1}, \ldots{}, D_{x_n})$ satisfying the \alldiff constraints of the problem $P$ it must hold that $e[\mathbf{d}] = e^1[\mathbf{d}] + \ldots{} + e^s[\mathbf{d}] \ge min^*(e^1) + \ldots{} + min^*(e^s) = min^*(e)$. 
Now if $min^*(e) > c$, then there are no $n$-tuples that satisfy both the \alldiff constraints of $P$ and the constraint $e \le c$. This proves the first part of the theorem. 

To prove the second part of the theorem, first notice that for any $n$-tuple $\mathbf{d}$ satisfying both the \alldiff constraints of the problem $P$ and the constraint $e \le c$ it holds that $min^*(e^1) + \ldots{} + min^*(e^{k-1}) + e^k[\mathbf{d}] + min^*(e^{k+1}) + \ldots{} + min^*(e^s) \le e^1[\mathbf{d}] + \ldots{} + e^{k-1}[\mathbf{d}] + e^k[\mathbf{d}] + e^{k+1}[\mathbf{d}] + \ldots{} + e^s[\mathbf{d}] \le c$. Therefore, $e^k[\mathbf{d}] \le c - (min^*(e) - min^*(e^k))$ for any solution of the problem $P$. This justifies the use of the value $b^k = c - (min^*(e) - min^*(e^k))$ as the upper bound for $e^k$. The rest of the proof is analogous to the proof of Theorem \ref{thm:main}, but $b^k$ is used instead of $c$. 
\qed

\section{Implementation} 
\label{sec:impl}

The algorithms described in Section \ref{sec:main} are implemented within our solver, called \texttt{argosmt} (\cite{bankovic_alldiff}). It is actually a DPLL($\mathcal{T}$)-based (\cite{dpll_t}) SMT solver (\cite{smt_in_sat_handbook}), consisting of a SAT solver based on so-called \emph{conflict-driven-clause-learning} (CDCL) algorithm (\cite{cdcl_in_sat_handbook}) and a set of decision procedures for theories of interest (known as \emph{theory solvers}). Beside some standard SMT theories, our \texttt{argosmt} solver also supports the \emph{CSP theory} which aims to formally describe the semantics of some of the
most frequently used global constraints --- notably the \alldiff constraint and the finite domain linear constraints.
The theory solver for the CSP theory consists of \emph{constraint handlers} and \emph{domain handlers}. For each constraint in the considered problem the theory solver instantiates one constraint handler which implements the appropriate filtering algorithm. Currently, the solver supports the filtering algorithms for the \alldiff constraints (\cite{regin}) and for the linear constraints (described in Section \ref{subsec:standard}). It also supports our improved filtering algorithm for the linear constraints described in Section \ref{subsec:improved}, as well as the algorithm given by Beldiceanu et al.~(\cite{beldiceanu}). The theory solver also contains one domain handler for each CSP variable. Its purpose is to maintain the state of the domain (e.g.~the minimum and the maximum value, the information about the excluded values, etc.).

The communication between the constraints is established by exchanging the \emph{domain restriction literals} via the \emph{assertion trail} of the SAT solver. Each domain restriction literal is of the form $x \Join a$, where $x$ is a CSP variable, $a$ is a value, and $\Join \in \{ =, \le, \ge, \ne, <, > \}$. These literals are used to logically represent prunings deduced by the filtering algorithms. The \emph{assertion trail} $M$ of the SAT solver is a stack-like structure that contains the literals that are currently set to \emph{true}. It enables efficient backtracking. When a filtering algorithm prunes some values from a variable's domain, it propagates the corresponding literals to the assertion trail. The SAT solver then notifies all other interested constraints about the new literals on the trail. Since the filtering algorithms for the linear constraints (Section \ref{sec:main}) reason about bounds, they always propagate 
inequality literals (i.e.~$x \le a$ or $x \ge a$). On the other hand, Regin's algorithm for the \alldiff constraints
(\cite{regin}) prunes individual values from the domains, so it propagates disequalities ($x \ne a$) and assignments ($x = a$). Finally, trivial implications concerning only one variable (such as $x \le 3 \Rightarrow x \ne 5$ or $x = 2 \Rightarrow x < 4$) are propagated by the corresponding domain handler. For efficiency, domain restriction literals are introduced to the SAT context lazily, first time when needed. 

In the context of an SMT solver, an important issue is the \emph{explaining} of the propagations and conflicts deduced by the theory solvers. An \emph{explanation} of a propagated literal $l \in M$  is a set $\{ l_1, \ldots{}, l_n \}$ of literals from $M$ that appear \emph{before} $l$ in $M$, such that $l_1, \ldots{}, l_n \Rightarrow l$. Similarly, an explanation of a conflict is a set $\{ l_1, \ldots{}, l_n \}$ of literals from $M$ such that $l_1, \ldots{}, l_n \Rightarrow \bot$.  Smaller explanations are preferred. In case of the theory solver for the CSP theory, each explanation must be generated by the handler that deduced the propagation (pruning) or the conflict (inconsistency) being explained. 

For a linear constraint, explaining of inconsistencies and prunings is reduced to explaining the bounds of the variables in the corresponding linear expression. Let us consider again the constraint $e \le c$, where 
$e \equiv a_1 \cdot x_1 + \ldots{} + a_n \cdot x_n$. If the inconsistency is detected, this means that $min(e) > c$ (or $min^*(e) > c$, in case of the improved algorithm). Since the value of $min(e)$ is determined by the values of $min(x_i)$ (or $max(x_i)$ in case of negative $a_i$), we must find the literals from $M$ that are responsible for establishing such bounds 
for the variables. This is done with assistance of the corresponding domain handler --- since it maintains the domain state, it has all the information about the removed values and the literals that caused their removals. This information is sufficient to explain the minimum or the maximum. For instance, if $min(x) = 2$ and the values below $2$ are pruned from the domain because of the literal $x \ge 2 \in M$, the domain handler will use $x \ge 2$ as an explanation for the minimum. On the other hand, if the literal $x \ne 2$ is added to $M$ later at some point, then the new minimum $min(x) = 3$ will be explained by $x \ge 2, x \ne 2$ (the first literal removed the values lower than $2$, and the second literal removed the value $2$ from the domain, making the minimum equal to $3$). In case the improved algorithm is used, the relevant \alldiff constraints should also be appended to the explanation, since they also affect the calculated value of $min^*(e)$. When explaining of prunings is concerned, in order to explain the bound for the variable $x_i$, it is sufficient to explain the minimum $min(e_{x_i})$ (or $min^*(e_{x_i})$) in a similar fashion.

\section{Experimental results}
\label{sec:experimental}

In this section we consider the experimental evaluation of our approach. The main goal of the evaluation is to compare the performance of the standard filtering algorithm (Algorithm \ref{algo:bounds_standard}) and our improved filtering algorithm (Algorithm \ref{algo:bounds_improved_gen}) within our solver.  Another goal is to compare our algorithm to the algorithm developed by Beldiceanu et al.~(\cite{beldiceanu}). This algorithm will be referred to as \emph{the bel algorithm} in the following text. This comparison will be possible only for the problems to which the \emph{bel} algorithm is applicable (since it requires all coefficients to be equal to $1$). Finally, the goal is also to compare our solver to the state-of-the-art constraint solvers. 

\paragraph{Solvers and problems.} In order to achieve these goals, the experiments were performed with the following solvers: 

\vspace{-3pt}
\begin{itemize}
 \item \texttt{argosmt-sbc} --- our solver\footnote{\nolinkurl{http://www.matf.bg.ac.rs/~milan/argosmt-5.5/}} using the standard filtering algorithm
 \item \texttt{argosmt-ibc} --- our solver using the improved filtering algorithm
 \item \texttt{argosmt-bel} --- our solver using the \emph{bel} algorithm
 \item \texttt{sugar}       --- Sugar\footnote{\nolinkurl{http://bach.istc.kobe-u.ac.jp/sugar/}} SAT-based constraint solver using \texttt{minisat}\footnote{\nolinkurl{http://http://minisat.se/}} as the back-end SAT solver
 \item \texttt{mistral} --- Mistral\footnote{\nolinkurl{http://homepages.laas.fr/ehebrard/mistral.html}} constraint solver
 \item \texttt{opturion} --- Opturion\footnote{\nolinkurl{http://www.opturion.com/}} \emph{lazy clause generation} solver (\cite{lcg})
\end{itemize}

\noindent{}We tested the solvers on the following five sets of instances:\footnote{Instances can be found at: \nolinkurl{http://www.math.rs/~milan/argosmt-5.5/instances.zip}}

\vspace{1ex}
\noindent{}\emph{kakuro} --- this set consists of 100 randomly generated kakuro instances
of size $30 \times 30$ (kakuro problem has been introduced in Example \ref{ex:kakuro}). All the instances are satisfiable. We used the standard encoding where each empty cell is represented by a variable with the domain $\{1,\ldots{},9\}$, and each line (horizontal or vertical) is constrained by a linear constraint (sum of variables in the line must be equal to the given number) and an \alldiff constraint (all variables in the line must be pairwise distinct). 

\vspace{1ex}
\noindent{}\emph{crypto} --- this set is inspired by the following famous
crypto-arithmetic puzzle: to each of 26 letters in English alphabet assign a \emph{value} --- a number from the set $\{1,\ldots{},26\}$ such that all letters have distinct values and the values of the letters in the words given bellow
sum to the given numbers:

\begin{center}
{\footnotesize
\begin{tabular}{lcllcllcllcl}
 ballet & = & 45, & cello & = & 43, & concert & = & 74, & flute & = & 30, \\
 fugue & = & 50, & glee & = & 66, & jazz & = & 58, & lyre & = & 47, \\
 oboe &  =  & 53, & opera & = & 65, & polka & = & 59, & quartet & = & 50, \\
 saxophone & = & 134, & scale & = & 51, & solo & = & 37, & song & = & 61, \\
 soprano & = & 82, & theme & = & 72, & violin & = & 100 & waltz & = & 34 \\
\end{tabular}
}
\end{center}
We randomly generated 100 similar satisfiable instances --- each instance consists of 20 randomly generated ``words'' of length between 4 and 9. Again, the encoding is straightforward: each letter is represented by a variable whose domain is the set 
$\{1,\ldots{},26\}$, there is one \alldiff constraint over all 26 variables (all letters must have distinct values), and for each word there is one linear constraint (the sum of corresponding variables must be equal to the given number). Notice that the coefficients in the linear constraints do not have to be equal to $1$, since a letter may appear more than once in a word. 

\vspace{1ex}
\noindent{}\emph{wqg} --- the third set consists of 100 randomly generated instances of the \emph{weighted quasigroup completion} problem (\cite{dichotomic}). The problem is similar to the standard quasigroup completion problem (\cite{gomes_phase_transition}), but in addition each cell $(i,j)$ of the corresponding \emph{latin square} has an assigned \emph{weight} $p_{ij}$ -- a positive integer from some predefined interval. The goal is to complete the quasigroup starting from the pre-given values, minimizing the value of $M = min_i(\sum_j p_{ij}x_{ij})$, where $x_{ij}$ are the values of the cells. Of course, this is an optimization problem, but we consider its decision variant --- is there a correct quasigroup completion such that $M \le K$, where $K$ is a positive number? The problem is encoded using \alldiff constraints over $x_{ij}$ variables (one \alldiff for each row and column). Also, for each row, we introduce a variable $y_i = \sum_j p_{ij}x_{ij}$ that represents the weighted sum for that row, and assert the clause $\bigvee_i (y_i \le K)$ --- it ensures that the minimum of the $y_i$ variables is at most $K$. All instances in the set are of the size $30\times 30$, the weights are between $1$ and $100$ and the boards are generated with around 42\% of cells filled in (the point of the phase transition (\cite{gomes_phase_transition})). The number $K$ is chosen based on some previously found quasigroup completion, so we know that all instances are satisfiable.

\vspace{1ex}
\noindent{}\emph{magic} --- the fourth set consists of 40 \emph{magic square} instances, downloaded from \emph{CSP-LIB}\footnote{\nolinkurl{http://www.csplib.org/Problems/prob019/data/MagicSquareCompletionInstances.zip}}. In general, the magic square problem may be described in the following way: the numbers $1, 2, \ldots{}, n^2$ should be placed in the cells of a $n\times n$ grid, such that all values in the grid are pairwise distinct and sums of each row, each column and each diagonal must be equal. Some values are already given, and the goal is to complete the grid in the described way. To encode the problem, we introduce a variable $x_{ij}$ for a cell in $i$th row and $j$th column of the grid. All such variables take values from the domain $\{1, \ldots{}, n^2 \}$. We have one \alldiff constraint that covers all $x_{ij}$ variables. Finally, we have one linear constraint for each row, each column and each diagonal, constraining the corresponding variables to sum up to the number $n \cdot (n^2 + 1) / 2$. Predefined values are simply encoded as equalities.  
This particular set of instances consists of 40 instances of size $9\times 9$, where 20 instances have 10 predefined values in the grid, and 20 have 50 predefined values in the grid.

\vspace{1ex}
\noindent{}\emph{gen-kakuro} --- the fifth considered problem is a generalization of the kakuro puzzle such that each empty cell in the grid has also a predefined \emph{weight} --- a positive number from $1$ to $9$. Now, each cell should be filled with a value from $1$ to $9$, such that the values in each line are pairwise distinct, and such that the values in each line  multiplied by the weights of the corresponding cells sum up to a given value. Compared to the standard kakuro, this time we have linear constraints with coefficients that are not all equal to $1$, which will help us to evaluate our algorithm better
in this general case. The set of instances consists of $100$ randomly generated satisfiable instances of size $100 \times 100$. 

\begin{figure}[!h]
\begin{center}
  \includegraphics[width=420pt]{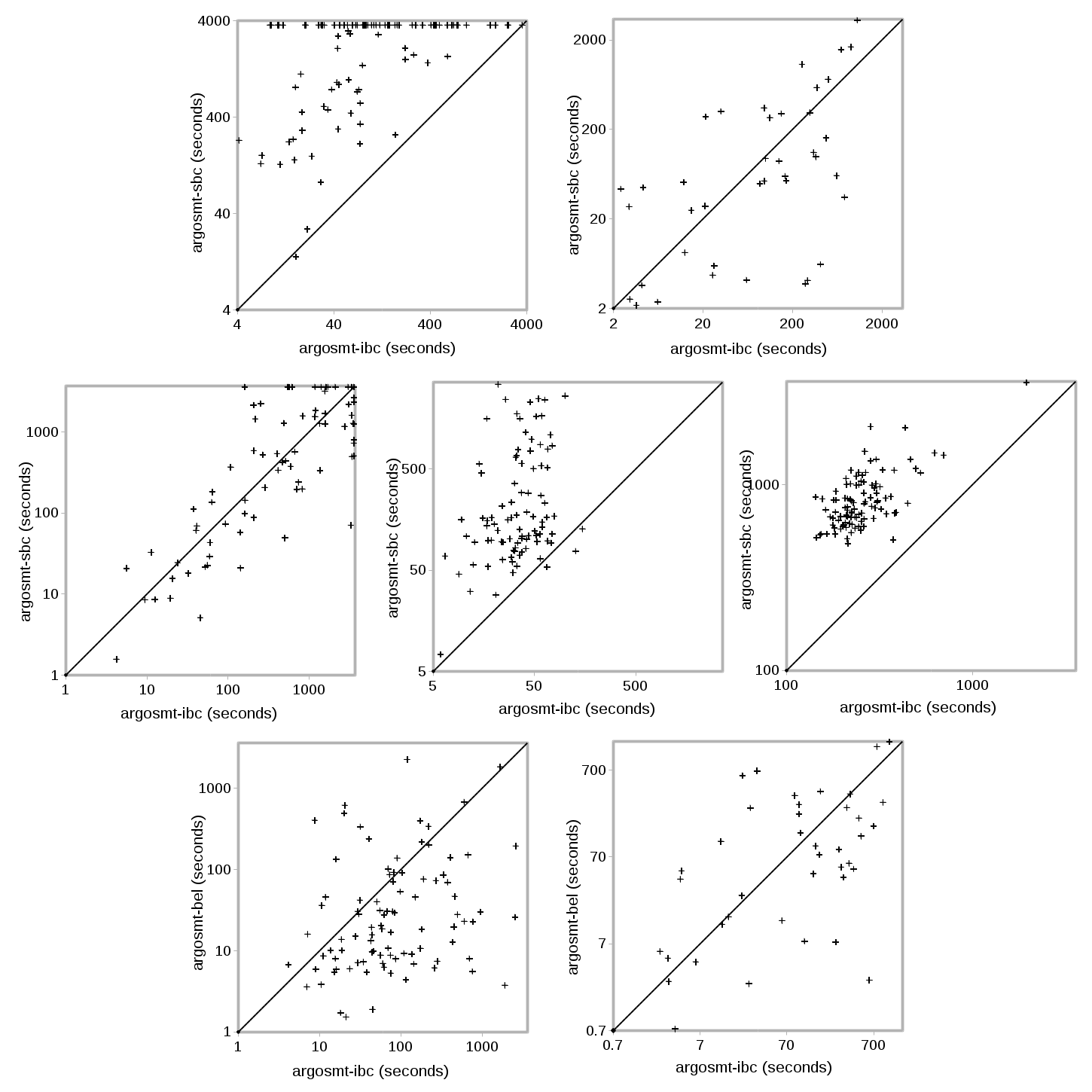}
\end{center}
\caption{Solving time comparison: \emph{kakuro} -- \emph{ibc} vs \emph{sbc} (top left), \emph{magic} -- \emph{ibc} vs \emph{sbc} (top right), \emph{crypto} --  \emph{ibc} vs \emph{sbc} (middle left), \emph{wqg} -- \emph{ibc} vs \emph{sbc} (center), \emph{gen-kakuro} -- \emph{ibc} vs \emph{sbc} (middle right), \emph{kakuro} -- \emph{ibc} vs \emph{bel} (bottom left), \emph{magic} -- \emph{ibc} vs \emph{bel} (bottom right)}
\label{fig:plot}
\end{figure}

\paragraph{Comparison with the standard algorithm.} Table \ref{tab:solved_avg} shows the numbers of solved instances and the average solving times for all solvers. For all instance sets, the cutoff time of 3600 seconds per instance is used (for unsolved instances, this cutoff time is used when the average time is calculated). The results show that the overall performance of our solver with the improved filtering algorithm is significantly better than with the standard filtering algorithm. Since the average time alone does not provide enough information about the runtime distribution, in Figure \ref{fig:plot} (the upper five charts) we show how runtimes relate on particular instances. The points above the diagonal line represent the instances for which the improved algorithm performed better than the standard algorithm. The charts look quite convincing for \emph{kakuro} instances (the top left chart), \emph{wqg} (the central chart), and \emph{gen-kakuro} (the middle right chart), where most of the points are above the diagonal. The improvement is not so obvious for \emph{magic} (top right) and \emph{crypto} (middle left) instances. 

\begin{table}[!h]
\begin{center}
{\footnotesize
\begin{tabular}{|c|c|c|c|c|}
 \cline{2-5}
 \multicolumn{1}{c|}{} & \multicolumn{2}{c|}{\textbf{kakuro}} & \multicolumn{2}{c|}{\textbf{magic}}  \\
 \cline{2-5}
 \multicolumn{1}{c|}{} & \textbf{\#instances} & \textbf{cutoff} & \textbf{\#instances} & \textbf{cutoff} \\
  \cline{2-5}
 \multicolumn{1}{c|}{} & 100 & 3600s & 40 & 3600s \\
 \cline{2-5}
 \multicolumn{1}{c|}{} & \textbf{\#solved} & \textbf{avg.~time} & \textbf{\#solved} & \textbf{avg.~time}  \\
 \hline 
 \textbf{argosmt-sbc} & 40 & 2521s & 40 & 296s \\
 \hline
 \textbf{argosmt-ibc} & 94 & 464s & 40 & 226s \\
 \hline 
 \textbf{argosmt-bel} & 99 & 196s & 40 & 198s \\
 \hline
 \textbf{sugar} & 100 & 91s & 40 & 57s \\
 \hline
 \textbf{mistral}  & 10 & 3305s & 39 & 111s \\
 \hline
 \textbf{opturion}  & 55 & 2126s & 40 & 10s  \\
 \hline
\end{tabular}
}
\end{center}

\begin{center}
{\footnotesize
\begin{tabular}{|c|c|c|c|c|c|c|}
 \cline{2-7}
 \multicolumn{1}{c|}{} & \multicolumn{2}{c|}{\textbf{crypto}} & \multicolumn{2}{c|}{\textbf{wqg}} & \multicolumn{2}{c|}{\textbf{gen-kakuro}} \\
 \cline{2-7}
 \multicolumn{1}{c|}{} & \textbf{\#instances} & \textbf{cutoff} & \textbf{\#instances} & \textbf{cutoff} & \textbf{\#instances} & \textbf{cutoff}   \\
  \cline{2-7}
 \multicolumn{1}{c|}{} & 100 & 3600s & 100 & 3600s & 100 & 3600s \\
 \cline{2-7}
 \multicolumn{1}{c|}{} & \textbf{\#solved} & \textbf{avg.~time} & \textbf{\#solved} & \textbf{avg.~time} & \textbf{\#solved} & \textbf{avg.~time} \\
 \hline 
 \textbf{argosmt-sbc} & 63 & 1745s & 92 & 722s & 99 &  904s \\
 \hline
 \textbf{argosmt-ibc}  & 70 & 1691s  & 100 & 44s & 100 & 283s \\
 \hline
 \textbf{sugar}  & 95 & 557s & 0 & 3600s & 100 & 42s \\
 \hline
 \textbf{mistral}  & 96 & 215s & 33 & 2412s & 0 & 3600s \\
 \hline
 \textbf{opturion}  & 84 & 985s & 100 & 43s & 100 & 62s \\
 \hline
\end{tabular}
}
\end{center}
\caption{The table shows for each solver and each instance set the number of solved instances within the given cutoff time per instance and the average solving time
         (for unsolved instances, the cutoff time is used)}
\label{tab:solved_avg}
\end{table}

Table \ref{tab:stats1} shows the information about the search space reduction. Since our solver is built on the top of a CDCL-based SAT solver, the best way to compare the explored search space is to look at the average numbers of decides and conflicts. For \emph{kakuro} instances, the search space is reduced almost four times. 
The situation is similar for \emph{gen-kakuro}, and is even more drastic for \emph{wqg} instances. The search space reduction is insignificant for \emph{crypto} and \emph{magic} instances. This mostly correlates with the average time reduction shown in Table \ref{tab:solved_avg}. Table \ref{tab:stats1} also shows the average portion of the execution time spent in the filtering algorithm. As expected, the improved algorithm does consume much more time (since its complexity is $O(n \log(n))$, compared to the linear complexity of the standard algorithm), but the reduction of the search space induced by our algorithm certainly justifies its usage.

\begin{table}[!h]
\begin{center}
{\footnotesize
\begin{tabular}{|c|c|c|c|c|c|c|}
 \cline{2-7}
 \multicolumn{1}{c|}{} & \multicolumn{3}{|c|}{\textbf{kakuro}} & \multicolumn{3}{|c|}{\textbf{magic}}  \\
 \cline{2-7}
 \multicolumn{1}{c|}{} & \textbf{sbc} & \textbf{ibc} & \textbf{bel} & \textbf{sbc} & \textbf{ibc} & \textbf{bel} \\
 \hline 
 \textbf{\#conflicts} & 274370 & 72785 & 45635 & 13206 & 11021 & 8516 \\
 \hline
 \textbf{\#decides} & 375863 & 98192 & 59064 & 25983 & 23126 & 17627 \\
 \hline
 \textbf{\%time in bc} & 9.1\% & 25.9\% & 24.8\% & 1.7\% & 5.6\% & 4.5\% \\
 \hline
\end{tabular}
}
\end{center}
\begin{center}
{\footnotesize
\begin{tabular}{|c|c|c|c|c|c|c|}
 \cline{2-7}
 \multicolumn{1}{c|}{} & \multicolumn{2}{|c|}{\textbf{crypto}} & \multicolumn{2}{|c|}{\textbf{wqg}}  & \multicolumn{2}{|c|}{\textbf{gen-kakuro}}  \\
 \cline{2-7}
 \multicolumn{1}{c|}{} & \textbf{sbc} & \textbf{ibc} & \textbf{sbc} & \textbf{ibc} & \textbf{sbc} & \textbf{ibc} \\
 \hline 
 \textbf{\#conflicts} & 187910 & 184409 & 31598 & 2345 & 12738 & 4464 \\
 \hline
 \textbf{\#decides} & 231503 & 225675 & 39852 & 6052 & 76492 & 23876 \\
 \hline
 \textbf{\%time in bc} & 14.6\% & 23.6\% & 5.0\% & 20.6\% & 1.7\% & 4.4\% \\
 \hline
\end{tabular}
}
\end{center}
\caption{The table shows the average numbers of conflicts and decisions and the average percents of time spent in the filtering algorithm}
\label{tab:stats1}
\end{table}

Table \ref{tab:stats2} shows the average percents of the bounds that are actually improved, compared to those obtained by the standard algorithm, and the average amounts of the improvements. 
We can see that, in case of \emph{kakuro} instances, 41\% of all calculated bounds are stronger when the improved algorithm is used. Although the average bound improvement looks small --- only 2.48, it is actually a significant improvement for \emph{kakuro} instances, since the domains of the variables are quite small (of size 9). In case of \emph{crypto} instances, the bound improvement is 1.44 on average, but the domains are much larger this time (of size 26), so this may be an explanation why the performance improvement on these instances is not so drastic as with \emph{kakuro} instances. The similar situation is for \emph{magic} instances, where the domains are of size 81, and the average improvement is only 1.56. On the other side, for \emph{wqg} instances, almost all bounds (95.5\%) are improved, and the average improvement is 1331, due to the large coefficients in the linear constraints. 

\begin{table}[!h]
\begin{center}
{\footnotesize
\begin{tabular}{|c|c|c|c|c|}
 \cline{2-5}
 \multicolumn{1}{c|}{} & \multicolumn{2}{|c|}{\textbf{kakuro}} & \multicolumn{2}{|c|}{\textbf{magic}}  \\
 \cline{2-5}
 \multicolumn{1}{c|}{} & \textbf{ibc/sbc} & \textbf{bel/sbc} & \textbf{ibc/sbc} & \textbf{bel/sbc} \\
 \hline 
 \textbf{\%bounds improved} & 41.0\% & 52.2\% & 24.0\% & 30.8\%  \\
 \hline 
 \textbf{avg. improvement} & 2.48 & 2.72  & 1.56 & 1.59  \\
 \hline 
\end{tabular}
}
\end{center}
\begin{center}
{\footnotesize
\begin{tabular}{|c|c|c|c|}
 \cline{2-4}
 \multicolumn{1}{c|}{} & \textbf{crypto} & \textbf{wqg}  & \textbf{gen-kakuro}  \\
 \cline{2-4}
 \multicolumn{1}{c|}{} & \textbf{ibc/sbc} & \textbf{ibc/sbc} & \textbf{ibc/sbc} \\
 \hline 
 \textbf{\%bounds improved} & 27.2\% & 95.5\% & 33.2\% \\
 \hline 
 \textbf{avg. improvement} & 1.44 & 1331.58 & 2.43 \\
 \hline 
\end{tabular}
}
\end{center}
\caption{The table shows the average percents of improved bounds and the average amounts of the bound improvements}
\label{tab:stats2}
\end{table}

\paragraph{Comparison with the \emph{bel} algorithm.} Let us now consider the comparison of our improved filtering algorithm (used by the \texttt{argosmt-ibc} solver) and the \emph{bel} algorithm (\cite{beldiceanu}), which is used by the \texttt{argosmt-bel} solver. 
The comparison is based on the two problems (\emph{kakuro} and \emph{magic}) to which both algorithms are applicable.
First, in Table \ref{tab:solved_avg} we can see that \texttt{argosmt-bel} is significantly better on \emph{kakuro}
instances, while there is no significant improvement for \emph{magic} instances. This is also confirmed in Figure 
\ref{fig:plot}, where the lower two charts represent the comparison of \texttt{argosmt-ibc} and \texttt{argosmt-bel} solvers (the bottom left chart is \emph{kakuro} and the bottom right chart is \emph{magic}). Again, the points above the diagonal line represent the instances for which \texttt{argosmt-ibc} is better. For \emph{kakuro} instances, most of the points are below the diagonal, which confirms that \texttt{argosmt-bel} is clearly better. This is not the case for \emph{magic} instances, where the points are almost evenly scattered on both sides of the diagonal. 

The search space reduction shown in Table \ref{tab:stats1} is also consistent with the previous observations. We can also see in Table \ref{tab:stats1} that the filtering algorithm in \texttt{argosmt-bel} takes a similar 
portion of time as in \texttt{argosmt-ibc}. This is expected, since both our algorithm and the \emph{bel} algorithm have the same time complexity (\cite{beldiceanu}).

Further information for comparison of the two algorithms is again given in Table \ref{tab:stats2}. For the solver \texttt{argosmt-bel}, the percent of improved bounds (compared to the standard algorithm, i.e.~the solver \texttt{argosmt-sbc}) for \emph{kakuro} instances is 52.2\% on average, which is somewhat better than our improved algorithm achieves (41\%). The similar situation is for \emph{magic} instances (30.8\% against 24\%), but the effect of this improvement is less significant on these instances because of the larger domains. The average improvement is similar for both algorithms. Further investigation reveals that, for \emph{kakuro}, about 34\% of conflicts detected by the \emph{bel} algorithm would not be detected by our improved algorithm (and that is less than 20\% in case of \emph{magic} instances).  We can conclude that, while the \emph{bel} algorithm will certainly perform better in practice than our algorithm, this difference in performance does not have to be drastic, despite the fact that the \emph{bel} algorithm enforces a stronger level of consistency. On the other side, an important advantage of our algorithm is its applicability to a much broader class of problems. 

\paragraph{Comparison with the state-of-the-art solvers.} Table \ref{tab:solved_avg} also shows that our solver is comparable to the state-of-the-art solvers used 
in the experiments. While our solver is not the best choice for solving any of the five problems (it is the second best choice for \emph{kakuro} and almost as good as \texttt{opturion} for \emph{wqg}), if we sum the results on all five instance sets, we can see that our solver (with the improved filtering algorithm) solves 404 instances in total which is the best result (\texttt{sugar} solves 335 instances, \texttt{mistral} solves 178 instances, and \texttt{opturion}
solves 379 instances in total). This means that our solver is the best choice on average for the five problems we have considered. 

\section{Related work} 
\label{sec:related}

When considered in isolation, there are well-known algorithms for establishing different forms of consistencies for both the \alldiff and the linear constraints. When the \alldiff constraint is concerned, the standard algorithm for establishing hyper-arc consistency is Regin's algorithm (\cite{regin}), while there are also several algorithms for establishing bound consistency (\cite{puget_alldiff_bounds}, \cite{mehlhorn_alldiff_bounds}, \cite{lopez_alldiff_bounds}). In case of the linear constraint, bound consistency may be established by the standard filtering algorithm described in Section \ref{subsec:standard}. It is quite straightforward and well-known in the literature (for instance, it is given in \cite{stuckey_bounds}).

There are also publications that consider combinations of the \alldiff and the linear constraints. One such approach is
given by Regin (\cite{regin_cost_gcc}), who developed an algorithm for establishing hyper-arc consistency on a \emph{global cardinality constraint with costs}. This quite general constraint can be used to model the conjunction $a_1 \cdot x_1 + \ldots{} + a_n \cdot x_n \le c \wedge \mathtt{alldifferent}(x_1, \ldots{}, x_n)$. There are two main drawbacks in this approach. First, we can only model the combination of one \alldiff and one linear constraint whose sets of variables must coincide. The second drawback is in the complexity of the algorithm, since it targets more general problem. Another interesting approach is 
given in the already mentioned work by Beldiceanu et al. (\cite{beldiceanu}). The authors present an algorithm that establishes bound consistency on the conjunction 
$x_1 + \ldots{} + x_n \le c\ \wedge\ \mathtt{alldifferent}(x_1, \ldots{}, x_n)$ and has $O(n \log(n))$ time complexity. The algorithm is, actually, more general, since the constraint $x_1 + \ldots + x_n \le c$ may be replaced, for instance,
by $x_1 \times \ldots{} \times x_n \le c$, or $x_1^2 + \ldots{} + x_n^2 \le c$. On the other hand, it does not permit arbitrary coefficients in the sum, like in the constraint $a_1 \cdot x_1 + \ldots{} + a_n \cdot x_n \le c$. Like the previous approach, this approach also requires that the set of variables appearing in the linear constraint coincides with the set of variables of the \alldiff constraint (or, at least, to be a subset of it). In the special case when all the coefficients in the sum are $1$, the algorithm developed by Beldiceanu et al.~(\cite{beldiceanu}) performs much better than Regin's algorithm (\cite{regin_cost_gcc}), despite the fact that only  bound consistency is established (\cite{beldiceanu}). Compared to these two approaches, our algorithm is more general, when combining the \alldiff and the linear constraints is concerned. First, it enables arbitrary coefficients (both positive or negative). Second, it does not limit us to use one \alldiff constraint in conjunction with 
the linear constraint and it does not require that the set of variables of the \alldiff coincides with the variables of the sum. In other words, we can have multiple \alldiff constraints that may only partially overlap with the variables of the linear constraint. The complexity of our algorithm is also $O(n \log(n))$, which makes it quite efficient. On the other hand, our algorithm does not establish bound (nor hyper-arc) consistency on a conjunction of an \alldiff and a linear constraint --- it just makes the pruning stronger to some extent, compared to the standard approach when both constraints are considered in isolation.

Another possible research path is to consider some modeling techniques that permit the constraint solvers to capture the interaction between the \alldiff and the linear constraints. The main idea is to add new implied linear constraints to the model imposing the bounds on sums of variables deduced from the \alldiff constraints. After that, other transformation techniques, such as \emph{common subexpression elimination} may be applied. For instance, assume the following set of constraints:
$x_1 + x_2 + x_3 + x_4 + x_5 + x_6 + x_7 \le 15$, $\mathtt{alldifferent}(x_1, x_2, x_3)$, $\mathtt{alldifferent}(x_5, x_6, x_7)$ (all variables have domains $\{1,\ldots{},10\}$). From the first \alldiff constraint, we can deduce that $x_1 + x_2 + x_3 \ge 6$ and $x_1 + x_2 + x_3 \le 27$ (the similar case is with the second \alldiff constraint). Furthermore, we can introduce new variables, $y = x_1 + x_2 + x_3$ and 
$z = x_5 + x_6 + x_7$ (whose domains are $\{ 6, \ldots{}, 27 \}$), and then rewrite the original constraint as 
$y + x_4 + z \le 15$. Using this extended model, the constraint propagation is made stronger. Indeed, in the original model, the propagator for the linear constraint can only deduce that $x_i \le 9$ (for $i \in \{1,\ldots{}, 7 \}$). In the extended model, the propagator for the rewritten linear constraint first deduces that $x_4 \le 3$, $y \le 8$, $z \le 8$. 
Then, from $y = x_1 + x_2 + x_3$ it can be deduced that $x_1 \le 6$, $x_2 \le 6$ and $x_3 \le 6$. One way to exploit this idea is to manually extend the model for some particular problem. For instance, in \cite{kakuro}, the special case of the mentioned Kakuro puzzle is considered. Similarly, in \cite{smith_walsh_golomb} and \cite{galinier_golomb}, the authors consider the application of such modeling techniques to the famous Golomb ruler problem. More generally, one can develop an algorithm for transforming models automatically, making the method applicable to a broader class of problems. For instance, in \cite{nightingale}, an algorithm for common subexpression extraction and elimination is described. The authors show that, using this transformation in conjunction with the linear constraints expressing the bounds deduced from the \alldiff constraints may significantly improve the results on \emph{Killer Sudoku} instances. The similar idea is employed in \cite{frisch_cgrass}, where a rule-based system for transforming models is presented. Compared to these modeling techniques, our algorithm may establish a stronger level of consistency. For instance, in the previous example, our algorithm (working on the original model) would deduce that $x_4 \le 3$, $x_1 \le 5$, $x_2 \le 5$ and $x_3 \le 5$. 

Finally, the idea of combining different types of global constraints in order to improve the constraint propagation is also 
employed in case of other constraints. For instance, in \cite{hnich_lex_sums}, the combination of the \emph{lexicographic ordering constraint} with two linear constraints is considered. On the other hand, in \cite{bessiere_alldiff_prec}, the authors consider a combination of an \alldiff constraint with the \emph{precedence constraints}.

\section{Conclusions}
\label{sec:conclusions}

In this paper we proposed an improved version of the standard filtering algorithm for the linear constraints
in CSP problems in the presence of the \alldiff constraints. The awareness
of the existence of the \alldiff constraints often permits the calculation of stronger bounds for variables which leads to more prunings. This may be useful for CSP problems that involve many \texttt{alldifferent}-constrained linear sums. 
We tested our approach on five
such problems and compared it to other relevant approaches as well as to some of the state-of-the-art solvers. The experiments confirmed the effectiveness of the improvement we proposed. The main advantage of our approach, compared
to other relevant approaches, is in its wider applicability, since it permits arbitrary coefficients in the linear 
expressions and may combine a linear constraint with multiple \alldiff constraints that partially overlap with its
variables. 

\section*{Acknowledgement}

This work was partially supported by the Serbian Ministry of Science grant 174021.
The author is grateful to Filip Mari\' c and to anonymous reviewers for very 
careful reading of the text and providing detailed and useful comments and remarks.

\bibliographystyle{alpha}
\bibliography{references}

\end{document}